\documentclass[a4paper,twoside,11pt]{article}
\usepackage[utf8]{inputenc}
\usepackage{amsmath}
\usepackage{amsfonts}
\usepackage{graphicx}
\usepackage{subeqnarray}
\usepackage{xcolor}
\usepackage[final]{pdfpages}
\usepackage{subcaption}
\usepackage[font=small,labelfont=bf]{caption}
\usepackage{geometry}
\usepackage{graphicx}
\usepackage[sort&compress,numbers]{natbib}
\usepackage{enumerate}
\usepackage{wasysym}
\usepackage[textsize=footnotesize]{todonotes}
\usepackage{url}

\usepackage{bm}

\definecolor{darkblue}{rgb}{0,0,0.6}
\definecolor{darkred}{rgb}{0.6,0,0}

\usepackage[colorlinks=true, urlcolor=black, citecolor=black, linkcolor=black, hyperfootnotes=false]{hyperref}

\usepackage{authblk}

\usepackage{natbib}
\usepackage{graphicx}

\newcommand{\IT}{\textrm{IT}}
\newcommand{\NT}{\textrm{NT}}
\newcommand{\MM}{\textrm{MM}}
\newcommand{\SK}{\textrm{SK}}

\newcommand{\short}{\textrm{short}}
\newcommand{\Long}{\textrm{long}}
\newcommand{\emp}{\textrm{emp}}
\newcommand{\F}{\textrm{F}}
\newcommand{\Prop}{\textrm{P}}

\newcommand{\avg}{\mathbb{E}}

\usepackage{authblk}

\author[1,2]{Michele Vodret\footnote{\url{mvodret@gmail.com}}}
\author[3]{Iacopo Mastromatteo}
\author[3]{Bence T\'oth}
\author[1,2,3]{Michael Benzaquen}
\affil[1]{Chair of Econophysics  Complex Systems, Ecole polytechnique, 91128 Palaiseau Cedex, France}
\affil[2]{Ladhyx, UMR CNRS 7646, Ecole polytechnique, 91128 Palaiseau Cedex, France}
\affil[3]{Capital Fund Management, 23-25, Rue de l’Université 75007 Paris, France}
\date{\today \vspace{-1cm}}
\title{Do fundamentals shape the price response? \\       A critical assessment of linear impact models}
\date{\today}
\begin{document}

\maketitle

\begin{abstract}
We compare the predictions of the stationary Kyle model, a microfounded multi-step linear price impact model in which market prices forecast  fundamentals through information encoded in the order flow, with those of the propagator model, a purely data-driven model in which trades mechanically impact prices with a time-decaying kernel.
We find  that, remarkably,  both models predict  the exact same price dynamics at high frequency, due to the emergence of universality at small time scales. On the other hand, we find those models to disagree on the overall strength of the impact function by a quantity that we are able to relate to the amount of excess-volatility in the market.
We reveal a crossover between a high-frequency regime in which the market reacts sub-linearly to the signed order flow, to a low-frequency regime in which prices respond linearly to order flow imbalances.
Overall, we reconcile results from the literature on market microstructure (sub-linearity in the price response to traded volumes)
with those relating to macroeconomically relevant timescales (in which a linear relation is typically assumed).
\end{abstract}
\providecommand{\keywords}[1]{\textbf{Keywords:} #1}


\tableofcontents

\setlength{\parskip}{\medskipamount}

\newpage

\section{Introduction}
Financial markets display very different dynamic properties depending  on the time window at which they are probed. If markets are observed in a small enough time window, where the price exhibits diffusive dynamics, then the slow evolution of fundamentals plays a minor role: in this regime  price dynamics is mostly driven by endogenous variables, such as past trades \cite{Gabaix}.  Conversely, by inspecting the market over larger time scales (e.g. months, years), fundamentals begin to play a more important role, and effects of mean-reversion to some notion of ``fair value'' start being visible~\cite{POTERBA198827,Blackok}.
Although the literature provides several phenomenological models capable of providing a multi-scale view of financial markets  \cite{Blackok,trendvalue}, no unique theoretical scenario is able to accommodate this crossover in a way that is both empirically accurate and economically sound. 

Seminal contributions in the market microstructure literature \cite{O'Harabook} include noisy rational expectation models~\cite{kyle,GLOSTEN198571} where the price partially reflects the underlying fundamental value of the asset,  providing an explanation of trade-induced price impact based on a price discovery mechanism. 
Unfortunately, some assumptions customarily made in this type of models make them unsuitable for calibration in real markets, where several aspects diverge from their idealized counterpart: in real markets  no fundamental price revelation is provided at any time and the order flow process exhibits long-range correlations~\cite{Lillo}. Some progress has been made in order to bridge the gap between microfounded price impact models and actual markets. In the Speculative Dynamics model (see \cite{Taub2}) the assumption of fundamental price revelation is dropped and a stationary price impact model is obtained. 
A further extension is provided by the stationary Kyle model~\cite{Vodret_2021}, which allows to consider arbitrary Gaussian signal and noise processes, de-facto including the empirically relevant case of strongly correlated order flow. In Ref.~\cite{Vodret_2021} it is also highlighted that the stationarity property allows to capture qualitatively the different dynamical regimes discussed above.

Interestingly, in the high frequency regime of the stationary Kyle model, where fundamentals are slowly varying and the price is diffusive, one recovers a linear equilibrium formally equivalent to that customarily described by the propagator model~\cite{Propagator} (see also Ref.~\cite{surprise}), which is an agnostic (i.e., not microfounded) model  able to provide a statistically accurate picture of financial markets at high frequency. 
Nevertheless, as we shall see in this paper, from a practical point of view, these two models are  very different. 
In fact, the excess-volatility puzzle \cite{Shiller} cannot be solved within the stationary Kyle model. This puzzle is instead avoided by the propagator model, which has no \emph{a-priori} on what the right price level should be, and  is thus allowed to set the level of price response by calibrating the model directly with the empirical price. Although the stationary Kyle model is not able to provide a solution to the excess-volatility puzzle, we shall argue that at small timescales, that has no effect on the qualitative price dynamics, because it only leads to an impact that is off by a multiplicative factor. Thus, the stationary Kyle model provides a microfoundation for the structure of the price impact shape, even though it misses a magnitude component that relates to the excess-volatility puzzle.

We discuss also the importance of the sampling scale at which markets are probed. It is  known from the propagator model literature, that if one studies financial markets at the trade by trade level,   microstructural effects related to order book details (such as selective liquidity taking) are of high relevance. In particular, it is well known \cite{Propagator} that a sub-linear price impact model gives higher predictive power than a linear one. We show that when analyzing coarse-grained data the opposite is true. 
In doing so, we provide a useful recipe to relate descriptions obtained with linear price impact models when the sampling scale is varied. 

The outline of the paper is the following. In section~2 we recall the basics of  
stationary Kyle model and the propagator model. In Section~3 we show how the stationary Kyle model captures the same shape of the price impact function as the propagator model, at high frequency. Section~4 highlights empirically the excess-volatility puzzle which affects the magnitude of the price impact function predicted by the stationary Kyle model and validates the linearity of price impact models if one considers coarse-grained data. Finally, in section~5 we conclude, suggesting an interesting way to reconcile the microfounded model with empirical price volatility.

\section{Linear models for price impact}

\subsection{The stationary Kyle model}
The stationary Kyle (sK) model~\cite{Vodret_2021} is a multi-step noisy rational expectations model with asymmetric information, where agents take actions at discrete steps $n$, and can trade a safe asset (\emph{cash}) and a risky asset~(\emph{stock}). 
A strategic, risk-neutral and forward-looking (with infinite horizon) agent (\emph{informed trader}, or IT)  possesses at each step $n$ privileged information about the  fundamental price $p^{\F}_n$ of the risky asset and exploits optimally this information by demanding a traded quantity~$q^{\IT}_n$ which maximizes his expected utility.
A passive agent (\emph{noise trader} or NT) accesses the market for exogenous reasons and trades in a  purely stochastic fashion. More precisely, his demand process~$q^\NT_n$ at step $n$ is a realization of a zero-mean, stationary Gaussian process with auto-covariance function (ACF) at lag $n$ given by~$\Omega^{\NT}_n$. Informed and noise traders are both modeled as liquidity takers. The third agent (\emph{market maker}, or  MM) is a liquidity provider who sets the transaction price $p^{\SK}_n$ in a risk-neutral and competitive way  observing only the realized excess demand $q_n = q^{\IT}_n+q^{\NT}_n$. 

The fundamental price is related to exogenous factors (e.g., dividends, earnings, market capitalisation). For the sake of argument, we suppose that the stock provides a stochastic payoff (dividend) $\mu_n$ to the owner at each step~$n$. Since we assume risk-neutrality of rational agents,  the presence of a fundamental price $p^{\F}_n$ can be justified as arising from the sum of future dividend cash flows:
\begin{equation}
    p^{\F}_n = \sum_{m \geq n} \mu_{m} .
\end{equation}
Dividends are modeled as realizations of an exogenous zero-mean, stationary Gaussian process.

Details about dynamics and information sets are given below. 
At the beginning of each time interval $(n-1,n]$ liquidity takers build demands for the risky asset, i.e., $q^i_n$ with $i \in \{\NT,\IT\}$. 
The informed trader's information  $\mathcal{I}^{\IT}_n$ is given by realized excess demand $q_n$, dividends $\mu_n$ and  NT's trades $q^{\NT}_n$ up  to step $n-1$ included. Given this information set, the IT  calculates an estimate of the fundamental price at each step $n$ as:
\begin{equation}
    p^{\IT}_n = \mathbb{E}[p^{\F}_n | \mathcal{I}^{\IT}_n] .
\end{equation}
 This estimate is the only new information about the fundamental price that is injected into the market at each  step $n$. The ACF of this process at lag $n$ is denoted by~$\Sigma^{\IT}_n$, which is Gaussian and stationary, due to the fact that dividends are realizations of a Gaussian stationary process. 
Based on his information set, at each step $n$, the IT chooses his optimal trading schedule maximizing his expected long-run gains and trades the first step of it, given by $q^\IT_n$. 

Then, the market maker clears the excess demand executing a trade $q^{\MM}_n = -q_n$ at transaction price $p^{\SK}_n$. The transaction price in a competitive market with risk-neutral traders is the optimal estimator of the fundamental price. The MM knows that the only new (unbiased) information about the fundamental price injected into the market at time $n$ is given by $p^\IT_n$, so the transaction price is equivalently the optimal estimator of it, given the MM's current information set  $\mathcal{I}^{\MM}_n$,  which coincides with realized excess demands up to time $n$ included.  In formula:
\begin{equation}
\label{eq:price}
    p^{\SK}_n =  \mathbb{E}[p^{\IT}_n|\mathcal{I}^{\MM}_n].
\end{equation}

Both the informed trader and the market maker  have exact knowledge about  statistical properties related to exogenous processes (dividends and NT's trades), as well as each other's strategy and information set structure.  Thus, rational agents are endowed with perfect structural knowledge.
Because of this assumption, the MM knows that the IT is injecting information in the market about fundamentals and he tries to extract it from the observed excess demand, in order to set a transaction price that accurately reflects the insider's  private information. The IT, on the other hand,  knows that the MM impounds into prices his filtered information about the fundamental price via a price impact function (see below), and decides how to trade while taking this into account. The MM sets the price in a competitive way, i.e., he doesn't earn money on average, while the NT accepts to lose money in favor of the IT. In change, the NT completes his trades at a transaction price that reflects optimally the underlying fundamental value.

The equilibrium arises endogenously as the result of agents' actions and is completely characterized by exogenous ACFs related to signal and noise (given by $\Sigma^{\IT}_n$ and  $\Omega^{\NT}_n$). 
The risk-neutrality assumption of rational agents and the restriction to Gaussian signal and noise implies an  equilibrium where the pricing rule $p^{\SK}_n$ is linear in realized excess demand. In formula:
\begin{equation}
    \label{eq:propagator_model}
    p^{\SK}_n = \sum_{m \leq n} G^{\SK}_{n-m}q_{m}.
\end{equation}
The price impact function $G^{\SK}_n$ can be found by numerically solving a functional equation that we investigated in Ref.~\cite{Vodret_2021}. The interpretation of $G^{\SK}_n$ that came out from that analysis is that the price impact function is the stationary Kalman filter related to the fundamental information injected in the market by the informed trader.
In what follows we show that, even though one has to resort to an iterative numerical procedure in order to compute $G^{\SK}_n$ directly from the definition of the equilibrium (see Ref.~\cite{Vodret_2021} for details), it is possible to derive the shape of it by exploiting two  equilibrium properties related to the price set by the MM and the excess demand.

At equilibrium, price and excess demand processes are Gaussian processes,  with zero mean and stationary ACF, given respectively by $\Sigma^{\SK}_n$ and $\Omega_n$. The temporal structure of ACFs related to excess demand and price set by the MM are  characterized by~\cite{Vodret_2021}:
\begin{subeqnarray}
\Omega_n 
\propto &\Omega^\NT_n + A\delta_n,
\slabel{eq:camouflage}
\\
\Sigma^{\SK}_n \propto& \Sigma^{\IT}_n .
\slabel{eq:price_efficiency}\label{eqscampri}
\end{subeqnarray}
The first property (Eq.~\eqref{eq:camouflage}) is linked to the IT's intention to hide his own trades in the excess demand in order to maximize his profits. In particular, the informed trader chooses his own optimal trading schedule such that the temporal structure of the excess demand ACF resembles that created by the NT alone, apart from an overall amplitude prefactor and a positive delta contribution at lag zero. The amplitude prefactor can be higher or smaller than one, depending on the degree of predictability of the NT's trade process. Because of the delta contribution, we called this property quasi-camouflage.  When the total order flow process exhibits some autocorrelation, perfect camouflage ($A \rightarrow 0$) is recovered in the regime where mean-reversion time scales of both fundamental price and NT's trades are large compared to the trading time-scale~\cite{Vodret_2021}. This is the empirically relevant case for sampling scales as large as  few hours. Note that obviously also a non-correlated NT's trade process ($\Omega^\NT_n \propto \delta_n$) implies perfect camouflage.
The second property (Eq.~\eqref{eq:price_efficiency}), that we called  price efficiency, shows that the price set by the MM  correctly reflects the statistical structure of the fundamental price process estimated by the IT  up to an amplitude factor. This amplitude factor is smaller than one because of the assumptions of  perfect structural knowledge, noisy environment and risk-neutrality of the MM. In fact, from Eq.~\eqref{eq:price}:
$
p^{\IT}_n = p^{\SK}_n+\epsilon_n,
$ where $\mathbb{E}[p^{\SK}_n \epsilon_n]=0$, due to the orthogonality property of optimal estimors. So, the following inequality holds:
$\Sigma^{\IT}_0 = \Sigma^{\SK}_0+\mathbb{E}[\epsilon_n^2] \geq \Sigma^{\SK}_0.
$ This means that the variance of the price set by the MM is smaller than that related to the IT's estimate of the fundamental price. In the empirical section (Subsec.~\ref{sub:excess_vol}) we give more details about this drawback of the model. 

Before moving on, let us show how one can use the two properties above to derive a simple equation that gives  the price impact function shape starting from the two input ACFs. From Eqs.~\eqref{eq:propagator_model} and \eqref{eqscampri} one can derive a relation between the ACF of excess demand,  $\Omega_n$, the ACF of the price,  $\Sigma^{\SK}_n$, and the impact function, $G^{\SK}_n$. In the case where perfect camouflage holds, this relation simplifies to:
\begin{equation}
\label{eq:combo}
\Sigma^{\IT}_n \propto \sum_{r\leq m} \sum_{s\leq m+n}  G^{\SK}_{m-r}G^{\SK}_{m+n-s} \Omega^{\NT}_{s-r}.
\end{equation}
For example, in the case of a Markovian market (which will be of interest in what follows) where $\Sigma^{\NT}_n \propto \delta_n$ and $\Sigma^{\IT}_n \propto \exp(- n/n^{\IT})$ one can see that $G^{\SK}_n\propto \exp(- n/n^{\IT})$.

\subsection{The propagator model}
Here we focus our attention on links and differences between Eq.~\eqref{eq:propagator_model}, and the equation that defines the propagator model~\cite{Propagator}. While the sK model is microfounded in terms of noisy heterogeneous agents where the price is the optimal estimator of the \emph{fundamental price} ($p^{\F}_n$), the phenomenological linear price impact model known as propagator model is meant to explain the \emph{empirical price} ($p^{\emp}_n$) in terms of past order flows. 
The equation that defines the linear propagator model is simply given by 
\begin{equation}
\label{eq:Prop}
 p^{\Prop}_n = \sum_{m\leq n} G^{\Prop}_{n-m} q_{m}.
\end{equation}
One finds that, besides its simplicity, the propagator model is able to explain a large fraction of price volatility, meaning that a large fraction of price moves is not related to exogenous shocks. 
In order to test the stationary Kyle model, in the following we confront it with the propagator model.

The calibration of the stationary Kyle and of the propagator model require different input processes.
The input processes needed for the calibration of the stationary Kyle model are given by ACFs related to signal and noise, i.e., $\Sigma^{\IT}_n$ and $\Omega^{\NT}_n$. On the other hand, the propagator model, which is not microfounded, is calibrated using  only publicly observable processes. In particular, the excess demand ACF  $\Omega_n$ and the response function $R^{\emp}_n  = \avg[q_{n} (p^{\emp}_{m+n} -p^{\emp}_{m-1})]$ are  needed~\cite{Propagator}.

 In the  regime where the price is diffusive, the shape of the price impact function is the same as that given by the sK model, since it solely depends on the total order flow ACF (see Sec.~\ref{sec:univ}). 
 The only difference in terms of prediction of these two models, if calibrated in a regime where the price is diffusive, is given by the magnitude of the price impact function. In particular, $G_0^{\Prop}\geq G_0^{\SK}$, since empirical prices are far more  volatile than fundamental values \cite{Shiller,summers} , as further discussed in Subsec~\ref{sub:excess_vol}.
 
 As a final remark, note that the propagator model proposed in Ref.~\cite{Propagator} is not implemented in the linear form given by Eq.~\eqref{eq:Prop}. In fact, the calibration in that work was made using trade by trade data, and  the predictive power  was maximized using a sub-linear function of signed volumes of past trades. In Subsec.~\ref{subsec:diffscales}, where aggregated data on sampling scale $\tau \geq 1$ minute are analyzed, we calibrate assuming the  propagator to be a linear function of signed volumes of past trades in Eq.~\eqref{eq:Prop}, showing how the predictive power increases with respect to a propagator model extremely concave in past total traded volumes (i.e., linear in the \emph{sign} of the trades).  In order to conduct this analysis it will be interesting to compare linear price impact models' outcomes when the sampling scale of the dataset varies. In the following we detail how one can relate predictions coming from calibrations conducted with data aggregated over different time windows.

\paragraph{Scaling relations as the sampling scale varies}
\label{sec:multi}

In order to relate the prediction of a discrete model with well-known time dependent quantities (such as the  mean-reversion time scale of the de-trended fundamental price) we need to specify the real-time counterpart $\tau$ of $1$ lag, i.e., the discretization time step (or the \emph{sampling scale}).  In practice, this timescale is set by the time resolution $\tau$ of the dataset at hand. For this reason it will be convenient to introduce a notation that makes this detail explicit. For example, if the price comes from a dataset with a one-minute resolution, $\tau=1$ minute, then we define $p_{n=1}^{(\tau)}$ as the price at time $1$ minute. This notation will be handy when  comparing  predictions coming from calibration performed on different coarse-grained version of the same dataset.
In the following we will be interested in comparing objects, like the price impact function, constructed  from processes defined at different sampling scales. In those cases the dependence on the sampling scale $\tau$ will be made explicit. 
If, in a given equation, no mention to a scale $\tau$ is made, it means that all the objects that appear are constructed with the same sampling scale $\tau$.

Suppose we have a dataset related to prices and total excess demand with a given sampling scale $\tau_\short$. We can calibrate the linear propagator model on it. Then, starting from the original dataset, one can construct a coarse-grained version of it with a sampling scale $\tau_{\Long} = r \tau_{\short}>\tau_{\short}$.
The relations between excess demand and price variables at different sampling scales are given by:
\begin{subeqnarray}	
    q^{(\tau_{\Long})}_n =& \sum_{m=(n-1)r+1}^{nr} q^{(\tau_{\short})}_m,
	\slabel{eq:order_f}   	
	\\
	p^{(\tau_{\Long})}_n =& p^{(\tau_{\short})}_{nr}
	\slabel{eq:price_f}	.
\end{subeqnarray}

Below we discuss the relations between total order flow ACF and response function calibrated at different sampling scales.
Taking into account Eq.~\eqref{eq:order_f}, a power law decay with exponent $\beta$  for the excess demand ACF (similar to the one empirically found in~\cite{Lillo}) and starting from the definition of excess demand auto-covariance, one obtains the following relation between total order flow ACFs at different sampling scales:  
\begin{equation}
\label{eq:scaling_order}
\Omega^{(\tau_\Long)}_n \approx \Omega^{(\tau_\short)}_{n} (\tau_\Long/\tau_\short)^{(2-\beta)}.
\end{equation}
In the diffusive regime, a similar calculation for the empirical response function gives:
\begin{equation}
\label{scaling_resp}
R^{(\tau_\Long)}_n \approx R^{(\tau_\short)}_{n} (\tau_\Long/\tau_\short)^{1/2+(2-\beta)/2}.
\end{equation}
In the following, for the sake of argument, we shall consider also total order flow processes following a non-correlated noise, instead of a strongly correlated one. Note that in this case the two equations above still holds, using $\beta=1$.
Due to the linearity of the price impact models we are working with, it is easy to understand the price impact function behavior when coarse-graining.
We find that (see App.~\ref{app:multiscale}) 
\begin{equation}
\label{eq:scaling_G}
G^{(\tau_{\Long})}_n \approx \frac{1}{r} \sum_{m=(n-1)r+1}^{nr} G^{(\tau_{\short})}_m,
\end{equation}
i.e., the price impact function on sampling scale $\tau_{\Long}$ is roughly the price impact function at sampling scale $\tau_\short$ averaged over a time window of length $\tau_{\Long}$.

\section{Price impact function in the stationary Kyle model}
\label{sec:univ}
\subsection{High- and low- frequency regimes}
\label{sub:freq}

The price efficiency property (Eq.~\eqref{eq:price_efficiency})  allows us to identify two different dynamical regimes  according to the mean-reversion time scale related to the  IT's estimate of the fundamental price, given by $\tau^{\F}$. In fact, the variogram  $V^{\SK}_n = \mathbb{E}[(p^{\SK}_{n+m}-p^{\SK}_m)^2]$ calculated with a sampling scale $\tau$ can be expressed (in non-pathological cases) as 
\begin{equation}
V^{\SK}_n = 2\Sigma^{\SK}_0\left( 1 - \frac{\Sigma^{\IT}_n}{\Sigma^{\IT}_0}\right) = \left\{
\begin{array}{ccc}
 \sigma^2 n  &  & n  \ll \tau^{\F}/\tau \\
 2\Sigma^\SK_0  &  & n  \gg \tau^\F/ \tau
\end{array}
\right. \, , \label{eq:diffusive_to_mr}
\end{equation}
for some positive $\sigma$, implying that the behavior of the price at time step $n$ is  diffusive when innovations in the fundamental price process are long-lived with respect to $n \tau$ and mean-reverting  in the opposite case.

These two regimes can be equivalently characterized by the response function  $R^{\SK}_n  = \avg[q_{n} (p^{\SK}_{m+n} -p^{\SK}_{m-1})]$, which is constant for  $ n \ll \tau^\F/\tau$ as a consequence of price diffusion, and decays for $n  \gg \tau^\F/\tau$ as a consequence of price mean reversion.  The model is thus consistent with the roughly flat empirical price response function reported in studies about high-frequency dynamics ~\cite{Propagator}  for $n \tau$ smaller than few days, since  $\tau^\F$ is of the order of several months/few years~\cite{Blackok}.

Understanding the behavior of the price impact function $G^{\SK}_n$ in the high-frequency regime  
is trickier, as its shape depends on the NT's order flow ACF. Let us consider two cases.
(a) If the noise trader's order flow is uncorrelated, the informed trader also trades in an uncorrelated fashion (due to the camouflage, Eq.~\eqref{eq:camouflage}). In this case the shape of the response and impact functions is identical in the high-frequency regime, because only a permanent price impact function  ensures diffusive prices in case of uncorrelated trade flow.
(b) Now, let us  consider  a noise trader's  order flow auto-correlated  over a time window given by $\tau^{\NT}$.  In this case, in order for the response function  to be flat at high frequency, the impact function has to be a decreasing function~\cite{Propagator}, meaning that the price response to the first trade anticipates the correlated flow in the same direction. The impact function will then relax to a non-zero quasi-permanent value, even though the flat behavior of the response function will be preserved thanks to a stream of correlated orders of length $\sim \tau^{\NT}$.  We call this impact quasi-permanent because it appears as a permanent one to any observer probing the market at scales $n  \ll \tau^\NT/\tau\ll \tau^\F/\tau$.
Obviously, at long times, i.e., $ n \gg \tau^\F/\tau \gg \tau^\NT/\tau$, all impact functions will decay to zero by construction due to the mean-reverting nature of the IT's estimate of the  fundamental price.

The qualitative picture in real markets at high frequency is very similar to case (b). In fact, the order flow  displays long-term correlations. More precisely, the order flow ACF is not integrable (auto-covariance is measured to slowly decay even across days). Because of this, in order to have diffusive prices and a flat response at high frequency, the impact function should slowly decay to zero in a precise non-integrable way in order to compensate for the persistence of the order flow \cite{Propagator}. 
Graphical details about cases (a) and (b) in the high and in the low frequency regime are provided in App.~\ref{app:examples}.

\subsection{Universality at high frequency}
\label{sub:univ}

The discussion above hints to the fact that a kind of universality emerges in the sK model: the specific dynamics of the fundamental price process  affects the short-term price dynamics  only through the diffusion constant $\sigma$, but does not shape the short-term impact function. 

\begin{figure}[t]
\centering
    \begin{subfigure}{0.45\textwidth}
        \includegraphics[scale=0.45]{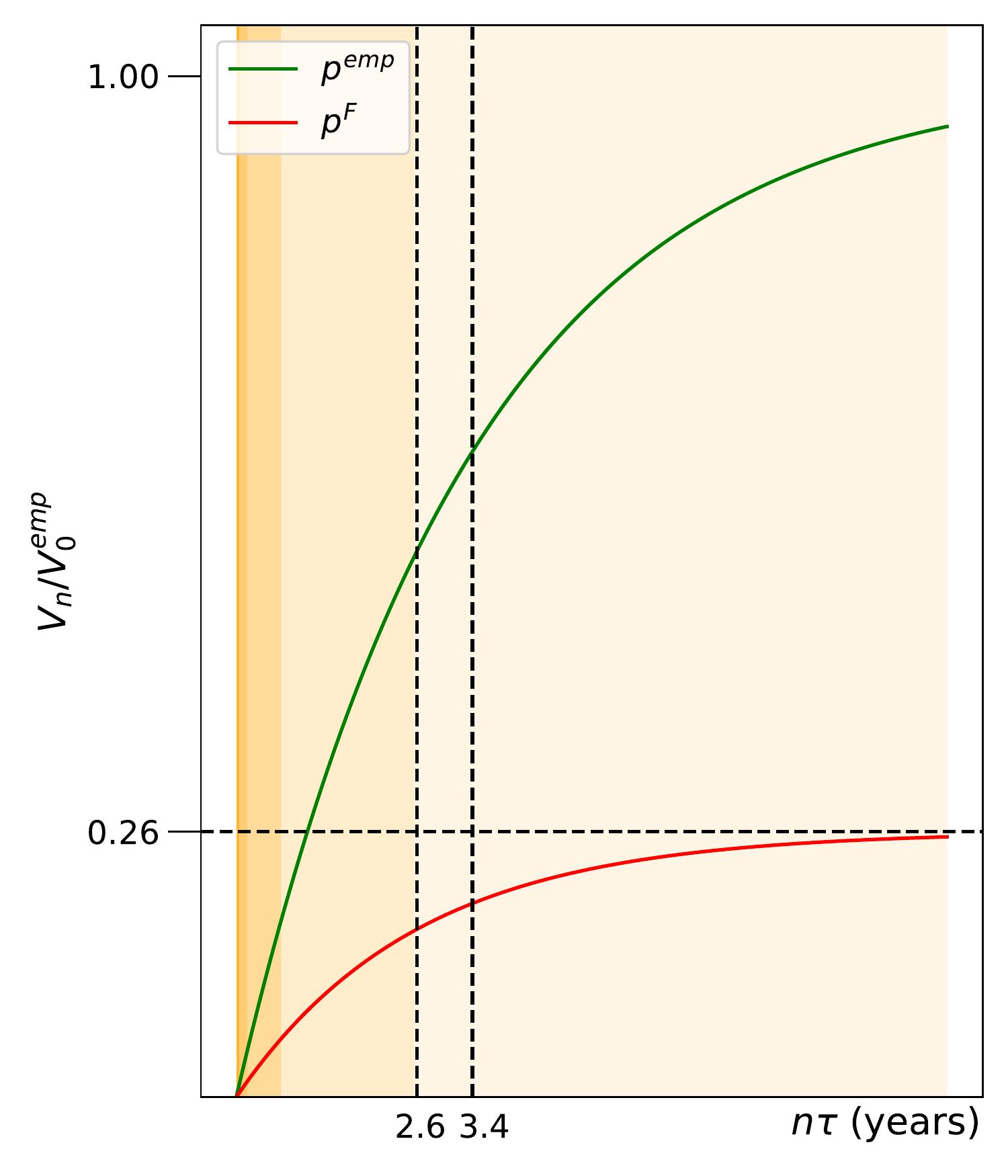}
    \end{subfigure}
\hfill
    \begin{subfigure}{0.45\textwidth}
        \includegraphics[scale=0.45]{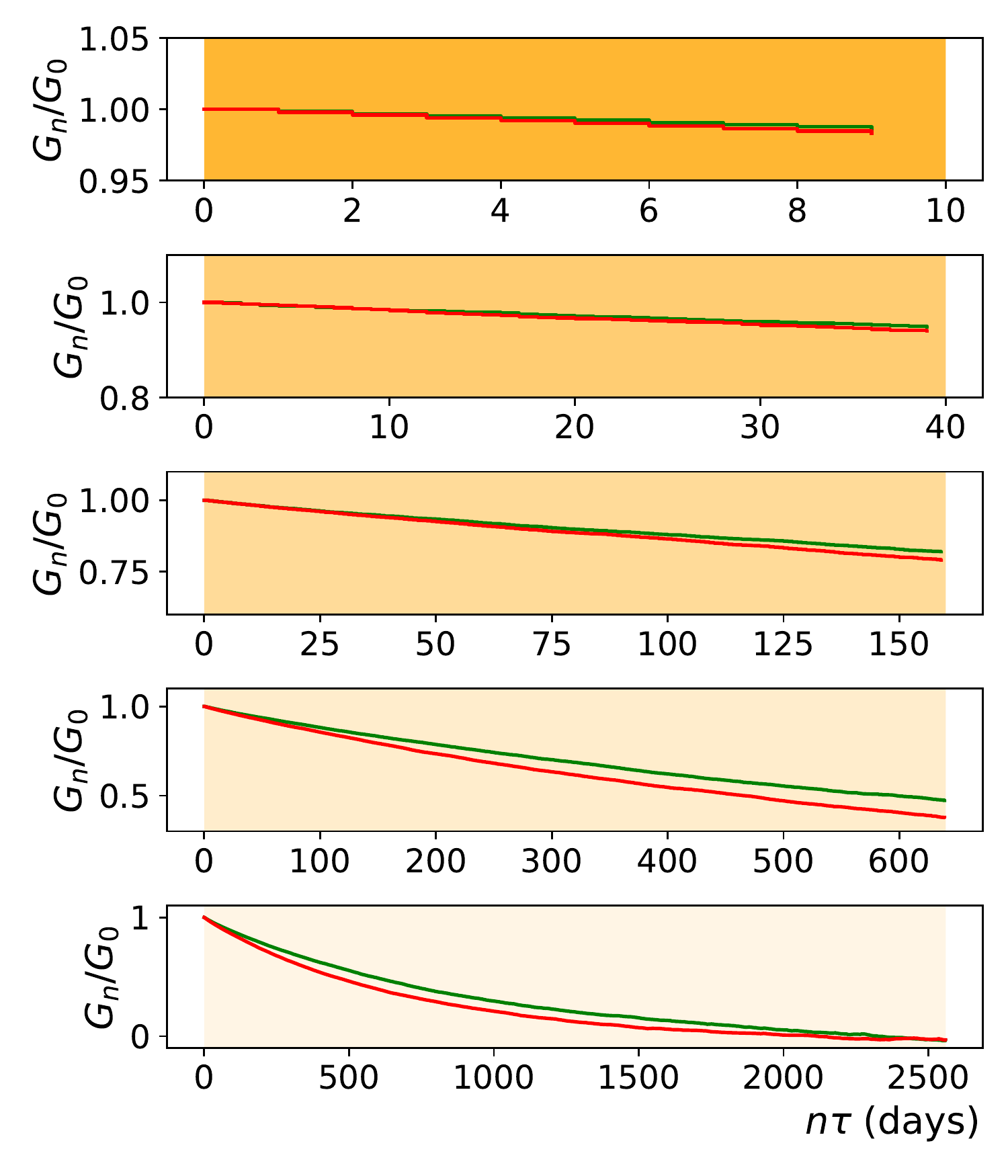}   
        \label{subfig:non_univ_G}
    \end{subfigure}
\caption{Universal time-dependence at high frequency and non-universal time-dependence at low frequency, for the case of a Markovian market with independent order flow. (Left) Variograms from synthetic data that comes from a  Markovian market with parameters given by Table 1. The different time windows used in the calibration (right panel) are highlighted by different shades of yellows.}
\label{Fig:non_universal}
\end{figure}

This statement is tested with synthetic data as follows. 
We generate two synthetic datasets, with sampling scale $\tau = 1$ day, that mimic Markovian price dynamics (with mean-reversion timescales and price volatility in line with those that will be presented in Table 1, i.e.,   $\tau^{\emp} = 250\times 3.4$ and $\tau^{\F} = 250\times 2.6$ days and $V^\F_\infty/V^\emp_\infty = 0.26$), while   order flow data are realizations of a non-correlated stochastic process. Variograms associated to these price processes are shown in the left panel of Figure~\ref{Fig:non_universal}. We then calibrate these data  with different time windows, graphically identified by different yellow bands: regions where the price undergoes pure diffusive dynamics are highlighted by an intense shade of yellow, and as the dynamics becomes affected by price mean reversion, the yellow band is lighter.
One can observe from the right panel of Figure~\ref{Fig:non_universal} that, modulo a global prefactor (absorbed by dividing the propagator function by its value at lag zero), the two price dynamics that we used ($p^{\F}$ and $p^{\emp}$) induce the same behavior for  $n\tau \ll \min\{\tau^{\emp},\tau^\F\}$ given approximately by a constant (or quasi-permanent) price impact function.
On the other hand, we see that as we approach a time window comparable with $\tau^{\emp}$ or $\tau^{\F}$, price impact functions start to drift away from each other, because they become sensitive to the different mean-reversion price dynamics.

Even though the collapse between price impact functions at high frequency, resulting from calibration made on different price processes but the same order flow process, takes place by construction (apart from an amplitude related to price volatility), the non-trivial thing that we are able to predict is up to which point the collapse holds, and relate it to effects linked to price mean-reversion  dynamics at low frequencies. 
Therefore,  the shape of the price impact function at high frequency is only affected by the shape of the order flow ACF, i.e.,  it is completely determined by the dynamics of flow anticipation. 

\section{Empirical results}
\label{sec:emp_res}

\subsection{Stationary Kyle model and excess-volatility}
\label{sub:excess_vol}

Our first empirical question is the following: how literally should the sK model be taken? More specifically, our construction relates the price impact kernel with properties of the fundamental price, predicting that even at high frequency the magnitude of trade-induced price jumps should be related to some notion of fundamental information that is propagated all the way down from low frequencies to high frequencies.
The stationary Kyle model implies in particular that it should be possible to deduce the price impact function from a proxy of the fundamental price of a stock (e.g., via dividends, earnings) and the properties of the signed order flow, without ever measuring the market price.

We tested this approach on a sample of monthly data (i.e., with  sampling scale  $\tau=1$ month),  related to empirical prices ($p^{\emp}$) and dividends (from which we constructed a proxy of the fundamental price $p^{\F}$) of the S$\&$P-500 index over $\sim 150$ years. The presentation of this dataset, the de-trending  and the calibration procedures are described in detail in App.~\ref{sub:low_freq_calib}. Note that  we  assume a Markovian dynamics for the fundamental price process (defined by a decay time scale $\tau^\F$ and an amplitude $\Sigma^\F_0$) and an independent order flow, which are a good approximation for such a large sampling scale $\tau$.

The left panel of Table~\ref{tab:time scales} contains the estimations for the mean-reversion time scale of the different prices. We find that the mean-reversion time scale of the fundamental price $\tau^{\F}$ is roughly in line with that deduced by the long-term behavior of the empirical market price  $\tau^{\emp}$, although the empirical market price seems to be slightly more persistent with respect to the fundamental price. Note that the price efficiency condition (Eq.~\eqref{eq:price_efficiency}) implies that the mean-reversion time scale of the price implied by the sK model $\tau^{\SK}$ is equal to that related to the fundamental price, i.e., $\tau^{\SK} = \tau^\F$. This means that the persistence amplification exhibited by the  empirical price  cannot be captured with the simple setting of the stationary Kyle model.

The second finding of this calibration is related to price volatility. The price volatility explained via the sK model is much smaller than that measured empirically as reported in the right panel of Table~\ref{tab:time scales}. We see that the squared fluctuations of both the fundamental price $\Sigma^{\F}_0$ and of the price set by the market maker $\Sigma^{\SK}_0$ in the sK model are much lower than the squared fluctuations of the empirical price $\Sigma^{\emp}_0$. This should come as no surprise in light of well-known results on excess-volatility~\cite{Shiller}. In fact the market price exhibits higher volatility with respect to what would be implied by any reasonable proxy for price fundamentals.
In our approach, the excess-volatility reported in that body of work is automatically inherited by the price impact function, due to the fact that the price is the optimal estimator of the fundamental price (see Eq.~\ref{eq:price}). Moreover, as explained above, the price set by the risk-neutral MM in a noisy environment always  reveals less information than the fundamental price so that we have the following chain of inequalities:  $\Sigma^{\SK}_0<\Sigma^{\F}_0<\Sigma^{\emp}$.

\begin{table}
  \begin{center}
    \begin{tabular}{c|c} 
      $\tau^{\emp}$ (years) & $\tau^{\F}$ (years)  \\
      \hline
      3.4 $\pm$ \  0.3 & 2.6 $\pm$ \ 0.3 \\
    \end{tabular}\hspace{2cm}\begin{tabular}{c|c} 

       $V^\F_\infty/V^{\emp}_\infty$ & $V^{\SK}_\infty/V^{\emp}_\infty$ \\
      \hline
       0.26 $\pm$ \  0.02 & 0.15 $\pm$ \ 0.03\\
    \end{tabular}
  \caption{Empirical results based on monthly recorded data about S$\&$P-500 index. (Left) Mean-reversion time scale of de-trended  empirical price, fundamental price and price set by the MM in the stationary Kyle setting. (Right) Price variance ratios as measures of explained empirical price volatility.}
  \label{tab:time scales}
  \end{center}
\end{table}

This last result shows that if one was to literally believe to the sK model, the price predictions that it implies would only account for a rather small portion of the volatility of the empirical market price. That is at odds with many empirical results concerning the predictive power of the propagator model, which is able to account for a very large portion (up to 60-70$\%$ if calibrated using trade by trade data) of the empirical market price variation. 
The source of this difference relies on the fact that the propagator model's input is the empirical price, and not the fundamental price as in the stationary Kyle model. In this way the propagator model is not affected by the excess-volatility puzzle, because there is not an explicit link to a notion of fundamental price. This implies that markets responds to the order flow as if it wasn't trying to anticipate the fundamental value of the risky asset, but something that would be much more in line with the market price.

\subsection{Predictive power of agnostic linear price impact models}
\label{subsec:diffscales}

Here we investigate the predictive power of the linear version of the propagator model given by Eq.~\eqref{eq:Prop}, showing that for sampling scale  $\tau \geq 1$ minute the results are satisfactory.  Our dataset contains order flow and price time-indexed data at the trade resolution for a variety of stocks  for a time range of 8 years (see details about data, de-trending and calibration procedures can be found in App.~\ref{sub:high_freq_calib}).

 \begin{figure}[t]
    \centering
    \includegraphics[scale = 0.55]{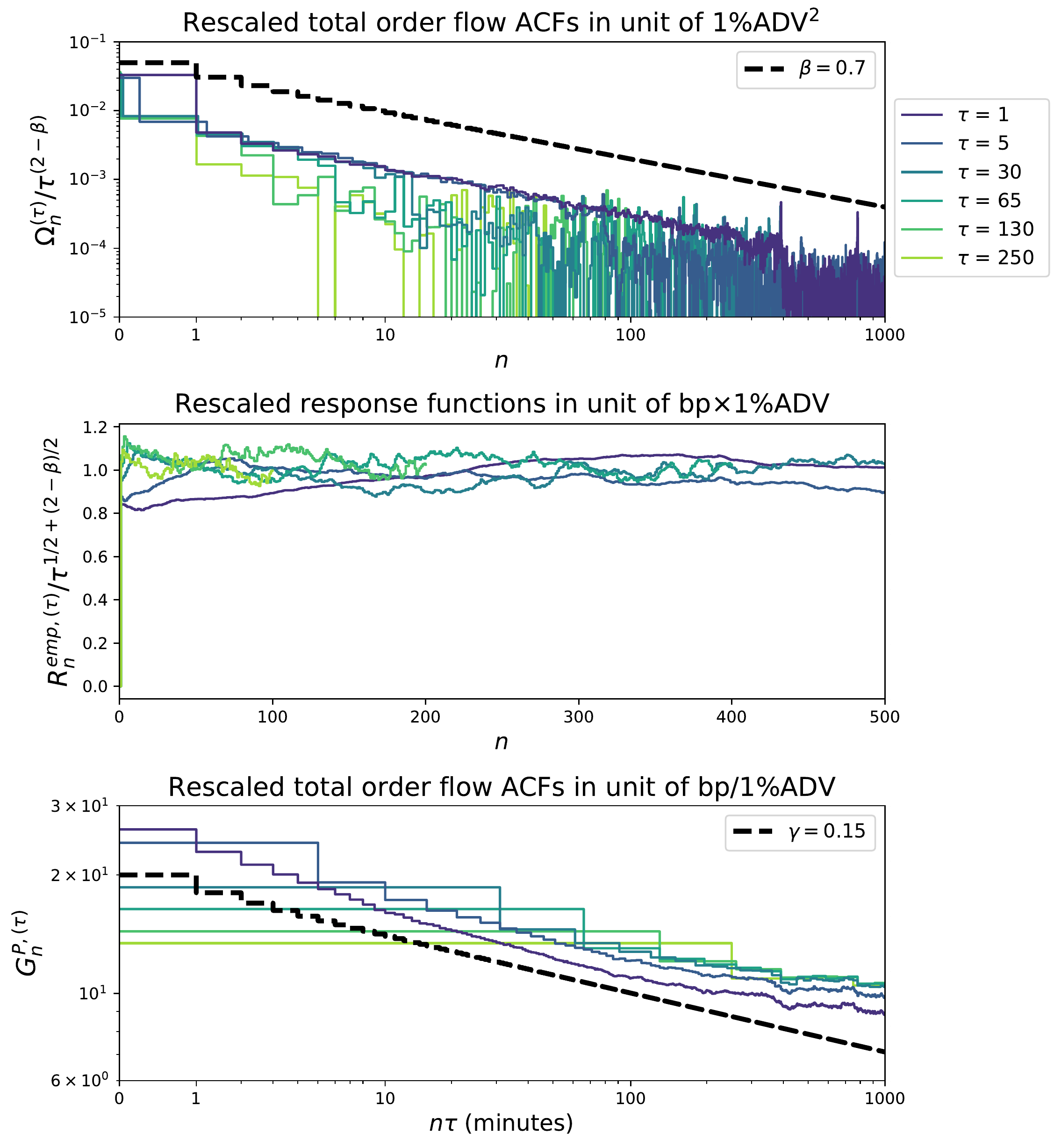}
    \caption{  Rescaled inputs and output of propagator model calibrations with different coarse-grained  high frequency data, averaged across stocks. Sampling scales $\tau$ related to different coarse-grained data are given in the legend, in units of minutes.  Volumes are measured in units of $1\%$ of average daily volume (ADV), while prices are measured in basis points (bp). (Top) Signed order flow ACFs, rescaled according to Eq.~\eqref{eq:scaling_order} and taking as a reference the total order flow ACF  with sampling scale $\tau = 1$ minute.   (Middle) Response functions, rescaled according to Eq.~\eqref{scaling_resp} and taking as a reference the response function with sampling scale $\tau = 1$ minute. (Bottom) Price impact functions. The dashed black lines in middle and bottom panel refers to decreasing power law function with exponents $\beta$ and $\gamma$, respectively, whose value are reported in the legends.}
    \label{fig:my_de-trended1}
\end{figure}

In Figure \ref{fig:my_de-trended1} we show the order flow covariance, the price response and the price impact function at sampling scales of 1, 5, 30, 65, 130 or 250 minutes. Note that for each  calibration, the analyzed time window is such that the price undergoes diffusive dynamics.
Order flow ACFs show a slow decay, compatible with a power law, with exponent lower than one. We find the power law exponent to be $\beta\approx0.7$. A
good collapse of the curves is obtained using the scaling given by Eq.~\eqref{eq:scaling_order}. 
The unit for the covariance is fraction of the average daily volume (ADV) squared. 
 As known from the literature, the price response has step-like shape, with a very quick increasing period and a plateau afterwards. This flattening out of the response is a sign of the diffusivity of the prices: no trivial future price prediction can be made from observing the trade flow. A good collapse of the curves is obtained using the scaling given by Eq.~\eqref{scaling_resp}.  The unit for the price response is basis points of the price times fraction of the average daily volume (ADV).
For the propagators at different sampling scales, we do not apply any rescaling, consistent with  Eq.~\eqref{eq:scaling_G}.  We see that, the propagator curves are comparable and quite similar at different sampling, validating the scaling found above.  Furthermore, the relation between the exponents related to power laws that fits the order flow ACF ($\beta$) and the propagator ($\gamma$) \cite{Propagator} holds, i.e., $\gamma =(1-\beta)/2 \approx~0.15$. The units are basis points of the price for a fraction of the average daily volume (ADV).

\begin{figure}[t]
\centering
\includegraphics[scale=0.5]{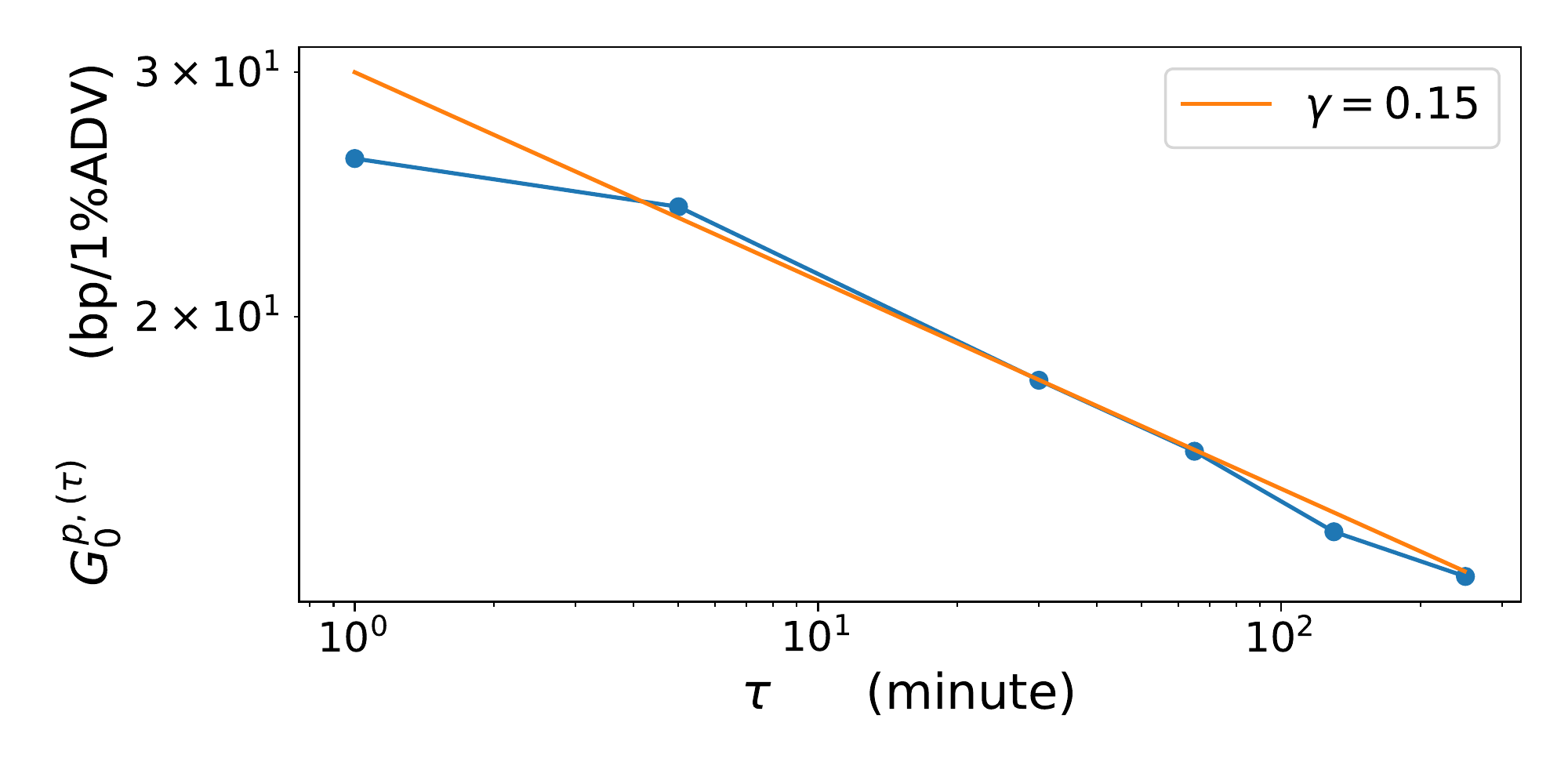}
\caption{Instantaneous price move in bp for a trade of 1$\%$ADV executed in a time-window of size $\tau$. }
\label{fig:G0}
\end{figure}

The price impact function at lag zero $G^{\Prop}_0$ suggests that a trade flow imbalance of $1\%$ of the average daily volume (ADV) leads to a price move of 26, 24, 18, 16, 14 or 13 basis points if executed in 1, 5, 30, 65, 130 or 250 minutes respectively. This decay of $G^{\Prop}_0$ for increasing $\tau$ is of importance. In fact, even though the total exchanged volume  dominates the overall price impact of a sequence of trades, the way in which the order flow is realized at the trade by trade level has a measurable effect, since the impact function is in fact decaying (albeit, at a small rate) and thus the concentration of trades plays a role.
The problem of optimal order execution (see, e.g., Refs.~\cite{Almgren_optimalexecution,Bertsimas,cartea2016a,Gatheral,Alfonsi}) deals precisely with the minimization of price impact induced costs under an exogenous constraint for the total size of the trade. 
  It is known in that context that it is possible to decrease transaction costs by exploiting the decay of the propagator function, and that such decrease is moderate whenever the decay of the impact function is slow. In case of a power-law decreasing $G_n \propto n^{-\gamma}$, the price impact related cost of an execution at constant rate $\phi$ during a period $T$ scales as $T^{-\gamma}$~\cite{Agent_based_model_pi}. This is explicitly shown in Fig.~\ref{fig:G0}, where  $G^{\Prop,(\tau)}_0$ is taken as a proxy for execution costs with different execution time ($\tau$ playing the role of $T$).

\begin{figure}[t]
    \centering
    \includegraphics[scale = 0.5]{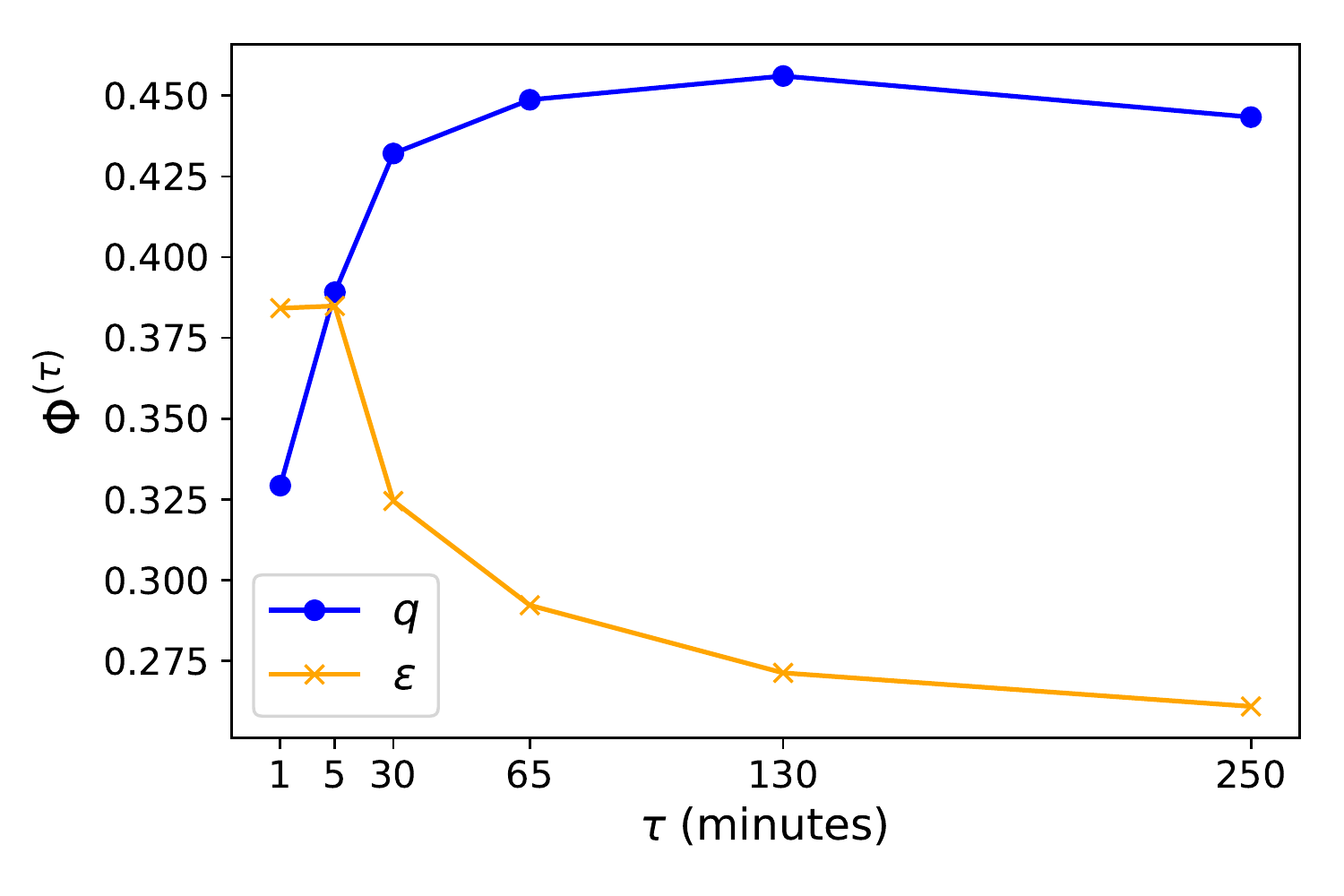}
    \caption{Ratio between predicted and empirical variograms at lag $\sim  4$ days,  as a function of the sampling scale $\tau$. Two linear propagator models are analyzed: price impact linear in  signed order flows (blue line) and linear in the sum of trades signs (orange line).}
    \label{fig:vargramratio}
\end{figure}

 In the remaining part of the section we study how much price  volatility can be explained by the linear version of the propagator model calibrated at different sampling scales. 
We define the ratio between predicted and empirical  price  variograms as \begin{equation}
\label{eq:var_ratio}
    \Phi^\Prop_n = \frac{V^\Prop_n}{V^\emp_n}.
\end{equation}
Figure ~\ref{fig:vargramratio} shows the metric $\Phi^{\Prop,(\tau)}$ at a lag corresponding to $\sim 4$ days. We show two curves.
The blue one ($q$) corresponds to a calibration using the sum of signed trade flow in the bin as before. The orange one ($\epsilon$) corresponds to a calibration using the sum of the signs of the trades in a bin.
Let's first concentrate on the blue curve ($q$). We find that the explained variance increases initially as the bin size increases, flattening out for longer sampling scales of a few tens of minutes. Most importantly, for all values of $\tau$ the explained variance is well above the ratio of $\Sigma^\F_0/\Sigma^{\emp}_0=0.26$ that can be found in Table \ref{tab:time scales}. This means that necessarily we have excess price response, which implies excess-volatility. This carries a very important message: the total trade flow imbalance is indeed of high relevance to explain  price moves when considering aggregated data. 
Let's now look at the orange curve ($\epsilon$), for which, interestingly,  one observes a better explanatory power than the $(q)$ model for short $\tau$ scales. Indeed, at the microstructural level, the actual traded volume is very much conditioned on the available liquidity in the limit order book, and the sign of the trade is more informative than the trade itself. In fact, it is rare that a trade penetrates more than one price level. This means that agents condition the size of their transactions on liquidity, making large transactions when liquidity is high and small transactions when it is low. Thus,  the information content is not related to the volume traded, but rather on the trade's sign. This effect undergoes the name of selective liquidity taking~\cite{bouchaud2008markets}.   When one considers aggregated data, however, the relation flips: as $\tau$ increases, the  information contained in $q$ becomes more valuable. An heuristic argument for this phenomenon is that the easier to manipulate a  market variable, the less it should carry information: manipulating $q$, the exchanged volume, means taking actual risk. For $\epsilon$, on the other hand, manipulation is much less risky: it can be done for example by placing many small trades in the market instead of few large trades. 

\section{Conclusions}

In this paper we compared two different linear price impact models very different in nature. The first one is a microfounded one (the stationary Kyle model), which explains why price impact arises based on an asymmetric information induced mechanism, and where the market price is the optimal estimator of the fundamental price. The second one is instead a phenomenological model (the propagator model), aimed at reconciling the diffusive price dynamics with the long-range correlation of order flows, which is able to capture a large fraction of market price volatility solely taking into account past trades.

With an empirical analysis we highlighted the impossibility to solve the excess-volatility puzzle in the stationary Kyle model. This implies that the price impact function calibrated with the stationary Kyle model will attain lower values than the one  obtained by calibrating the propagator model. Nevertheless, the shape of the two obtained price impact functions is the same at high frequency, where the price is diffusive. This is explained by the microfounded model narrative by saying that the slow evolution of fundamentals does not shape the high frequency dynamics of the price process, but only affects the magnitude of the price impact function via the diffusion constant related to the price.

We stress-tested the linearity assumption of the analyzed price impact models, demonstrating that the propagator model has a high predictive power if the sampling scale is larger than the typical trading timescale, because effects related to order book dynamics (such as selective liquidity taking)  can be neglected. In order to check the robustness of our findings we applied scaling arguments to obtain relationships between price impact functions at different sampling scales.

Note that the excess-volatility puzzle can be solved in a setting with asymmetric information, like the stationary Kyle model, if assumptions such as risk-neutrality and/or perfect structural knowledge are relaxed. The accuracy of the microfounded model will  increase if, for example, risk-aversion \cite{Subrahmanyam,GuoKyle} and/or learning dynamics \cite{Timmermann2,Timmermann,Hommes,Branch} are introduced. We look forward to explore this research direction.

\section*{Acknowledgments}

We thank J.-P. Bouchaud for many insightful discussions on these topics. This research was conducted within the \emph{Econophysics \& Complex Systems} Research Chair, under the aegis of the Fondation du Risque, the Fondation de l'Ecole polytechnique, the Ecole polytechnique and Capital Fund Management.

\bibliography{references}

\begin{thebibliography}{27}
\providecommand{\natexlab}[1]{#1}
\providecommand{\url}[1]{\texttt{#1}}
\expandafter\ifx\csname urlstyle\endcsname\relax
  \providecommand{\doi}[1]{doi: #1}\else
  \providecommand{\doi}{doi: \begingroup \urlstyle{rm}\Url}\fi

\bibitem[Gabaix and Koijen(2021)]{Gabaix}
X.~Gabaix and R.~S.~J. Koijen.
\newblock In search of the origins of financial fluctuations: The inelastic
  markets hypothesis.
\newblock Working Paper 28967, National Bureau of Economic Research, 2021.

\bibitem[Poterba and Summers(1988)]{POTERBA198827}
J.~M. Poterba and L.~H. Summers.
\newblock Mean reversion in stock prices: Evidence and implications.
\newblock \emph{Journal of Financial Economics}, 22\penalty0 (1):\penalty0
  27--59, 1988.

\bibitem[Bouchaud et~al.(2017)Bouchaud, Ciliberti, Lemperiere, Majewski,
  Seager, and Ronia]{Blackok}
J.-P. Bouchaud, S.~Ciliberti, Y.~Lemperiere, A.~Majewski, P.~Seager, and
  K.~Ronia.
\newblock Black was right: Price is within a factor 2 of value.
\newblock \emph{SSRN Electronic Journal}, 11 2017.

\bibitem[Majewski et~al.(2020)Majewski, Ciliberti, and Bouchaud]{trendvalue}
A.~Majewski, S.~Ciliberti, and J.-P. Bouchaud.
\newblock {Co-existence of trend and value in financial markets: Estimating an
  extended Chiarella model}.
\newblock \emph{Journal of Economic Dynamics and Control}, 112\penalty0 (C),
  2020.

\bibitem[O'Hara(1998)]{O'Harabook}
M.~O'Hara.
\newblock \emph{Market Microstructure Theory}.
\newblock Blackwell Publishing Ltd, 1998.

\bibitem[Kyle(1985)]{kyle}
A.~Kyle.
\newblock Continuous auctions and insider trading.
\newblock \emph{Econometrica}, 53\penalty0 (6):\penalty0 1315--35, 1985.

\bibitem[Glosten and Milgrom(1985)]{GLOSTEN198571}
L.~R. Glosten and P.~R. Milgrom.
\newblock Bid, ask and transaction prices in a specialist market with
  heterogeneously informed traders.
\newblock \emph{Journal of Financial Economics}, 14\penalty0 (1):\penalty0
  71--100, 1985.

\bibitem[Lillo and Farmer(2004)]{Lillo}
F.~Lillo and J.~Farmer.
\newblock The long memory of the efficient market.
\newblock \emph{Studies in Nonlinear Dynamics Econometrics}, 8\penalty0
  (3):\penalty0 1--35, 2004.

\bibitem[Bernhardt et~al.(2010)Bernhardt, Seiler, and Taub]{Taub2}
D.~Bernhardt, P.~Seiler, and B.~Taub.
\newblock Speculative dynamics.
\newblock \emph{Economic Theory}, 44\penalty0 (1):\penalty0 1--52, 2010.

\bibitem[Vodret et~al.(2021)Vodret, Mastromatteo, T{\'{o}}th, and
  Benzaquen]{Vodret_2021}
M.~Vodret, I.~Mastromatteo, B.~T{\'{o}}th, and M.~Benzaquen.
\newblock A stationary kyle setup: microfounding propagator models.
\newblock \emph{Journal of Statistical Mechanics: Theory and Experiment},
  2021\penalty0 (3):\penalty0 033410, 2021.

\bibitem[Bouchaud et~al.(2004)Bouchaud, Gefen, Potters, and Wyart]{Propagator}
J.-P. Bouchaud, Y.~Gefen, M.~Potters, and M.~Wyart.
\newblock Fluctuations and response in financial markets: the subtle nature of
  ‘random’ price changes.
\newblock \emph{Quantitative Finance}, 4\penalty0 (2):\penalty0 176--190, 2004.

\bibitem[Farmer et~al.(2006)Farmer, Gerig, Lillo, and Mike]{surprise}
J.~D. Farmer, A.~Gerig, F.~Lillo, and S.~Mike.
\newblock Market efficiency and the long-memory of supply and demand: is price
  impact variable and permanent or fixed and temporary?
\newblock \emph{Quantitative Finance}, 6\penalty0 (2):\penalty0 107--112, 2006.

\bibitem[Shiller(1981)]{Shiller}
R.~Shiller.
\newblock Do stock prices move too much to be justified by subsequent changes
  in dividends?
\newblock \emph{American Economic Review}, 71:\penalty0 421--36, 01 1981.

\bibitem[Summers(1986)]{summers}
L.~H. Summers.
\newblock Does the stock market rationally reflect fundamental values?
\newblock \emph{The Journal of Finance}, 41\penalty0 (3):\penalty0 591--601,
  1986.

\bibitem[Almgren and Chriss(2001)]{Almgren_optimalexecution}
R.~Almgren and N.~Chriss.
\newblock Optimal execution of portfolio transactions.
\newblock \emph{Journal of Risk}, pages 5--39, 2001.

\bibitem[Bertsimas and Lo(1998)]{Bertsimas}
D.~Bertsimas and A.~Lo.
\newblock Optimal control of execution costs.
\newblock \emph{Journal of Financial Markets}, 1\penalty0 (1):\penalty0 1--50,
  1998.

\bibitem[Cartea and Jaimungal(2016)]{cartea2016a}
A.~Cartea and S.~Jaimungal.
\newblock Incorporating order-flow into optimal execution.
\newblock \emph{Mathematics and Financial Economics}, 10\penalty0 (3):\penalty0
  339–364, 2016.

\bibitem[Gatheral(2010)]{Gatheral}
J.~Gatheral.
\newblock No-dynamic-arbitrage and market impact.
\newblock \emph{Quantitative Finance}, 10\penalty0 (7):\penalty0 749--759,
  2010.

\bibitem[Alfonsi et~al.(2010)Alfonsi, Fruth, and Schied]{Alfonsi}
A.~Alfonsi, A.~Fruth, and A.~Schied.
\newblock Optimal execution strategies in limit order books with general shape
  functions.
\newblock \emph{Quantitative Finance}, 10\penalty0 (2):\penalty0 143--157,
  2010.

\bibitem[Mastromatteo et~al.(2014)Mastromatteo, T\'oth, and
  Bouchaud]{Agent_based_model_pi}
I.~Mastromatteo, B.~T\'oth, and J.-P. Bouchaud.
\newblock Agent-based models for latent liquidity and concave price impact.
\newblock \emph{Phys. Rev. E}, 89:\penalty0 042805, 2014.

\bibitem[Bouchaud et~al.(2008)Bouchaud, Farmer, and Lillo]{bouchaud2008markets}
J.-P. Bouchaud, J.~D. Farmer, and F.~Lillo.
\newblock How markets slowly digest changes in supply and demand.
\newblock \emph{Handbook of Financial Markets: Dynamics and Evolution}, 2008.

\bibitem[Subrahmanyam(1991)]{Subrahmanyam}
A.~Subrahmanyam.
\newblock {Risk Aversion, Market Liquidity, and Price Efficiency}.
\newblock \emph{Review of Financial Studies}, 4\penalty0 (3):\penalty0
  416--441, 1991.

\bibitem[Guo and Kyle(2005)]{GuoKyle}
M.~Guo and A.~Kyle.
\newblock Dynamic strategic informed trading with risk-averse market makers.
\newblock Working Paper, Duke University, 2005.

\bibitem[Timmermann(1993)]{Timmermann2}
A.~G. Timmermann.
\newblock How learning in financial markets generates excess volatility and
  predictability in stock prices.
\newblock \emph{The Quarterly Journal of Economics}, 108\penalty0 (4):\penalty0
  1135--1145, 1993.

\bibitem[Timmermann(1996)]{Timmermann}
A.~G. Timmermann.
\newblock {Excess Volatility and Predictability of Stock Prices in
  Autoregressive Dividend Models with Learning}.
\newblock \emph{The Review of Economic Studies}, 63\penalty0 (4):\penalty0
  523--557, 10 1996.

\bibitem[Hommes and Zhu(2014)]{Hommes}
C.~Hommes and M.~Zhu.
\newblock Behavioral learning equilibria.
\newblock \emph{Journal of Economic Theory}, 150:\penalty0 778--814, 2014.

\bibitem[Branch(2016)]{Branch}
W.~A. Branch.
\newblock {Imperfect knowledge, liquidity and bubbles}.
\newblock \emph{Journal of Economic Dynamics and Control}, 62\penalty0
  (C):\penalty0 17--42, 2016.

\end{thebibliography}

\newpage
\appendix

\section{Price impact function scaling varying the  sampling scale}
\label{app:multiscale}

Consider a slowly varying kernel $G_n^{(\tau_{\short})}$ (assumption which will be realized empirically at small enough sampling scales $\tau_\short$). One can `zoom-out' in time, by defining a new coarse-grained model with sampling scale $\tau_{\Long} = r \tau_{\short}>\tau_{\short}$.  We show how the impact function changes when the sampling scale is changed. The argument goes as follows:
\begin{eqnarray*}
\label{eq:multiscale_prop}
p^{(\tau_{\short})}_{nr}
&=& 
\sum_{m \leq n}
\left(
G^{(\tau_{\short})}_{(n - m)r} q^{(\tau_{\short})}_{mr} + G^{(\tau_{\short})}_{(n - m)r -1} q^{(\tau_{\short})}_{mr-1} +\dots + G^{(\tau_{\short})}_{(n - m)r - r + 1} q^{(\tau_{\short})}_{mr - r + 1}
\right) \\
&\approx&
\sum_{m \leq n}
\underbrace{\frac{1}{r}(
G^{(\tau_{\short})}_{(n - m)r } + G^{(\tau_{\short})}_{(n - m)r -1} + \dots + G^{(\tau_{\short})}_{(n - m)r  - r + 1})
}_{ G^{(\tau_{\Long})}_{n-m} }
(
\underbrace{
 q^{(\tau_{\short})}_{mr} + q^{(\tau_{\short})}_{mr-1} + \dots+ q^{(\tau_{\short})}_{mr - r + 1}
 }_{q^{(\tau_{\Long})}_{m}}
) 
\\
&=& p^{(\tau_{\Long})}_n.
\end{eqnarray*}
Equations above indicate that when downsampling the model to scale $\tau_{\Long}$, the price is linear -at a first approximation-  in the total imbalance over bins of size equal to  $\tau_{\Long}$. The price impact function at scale $\tau_{\Long}$ is the arithmetic mean over a time window equal to $\tau_{\Long}$ of the price impact function defined at scale $\tau_{\short}$. Even though one has exact equality only in the case of a perfectly constant kernel or a perfectly constant order flow imbalance, the approximation will still retain a good explanatory power as long as $G^{(\tau_{\short})}_n$ is smooth enough.

\section{Linear price impact functions from different price dynamics}
\label{app:examples}

\begin{figure}
    \begin{subfigure}{1\textwidth}
    \centering
    \includegraphics[scale = 0.5]{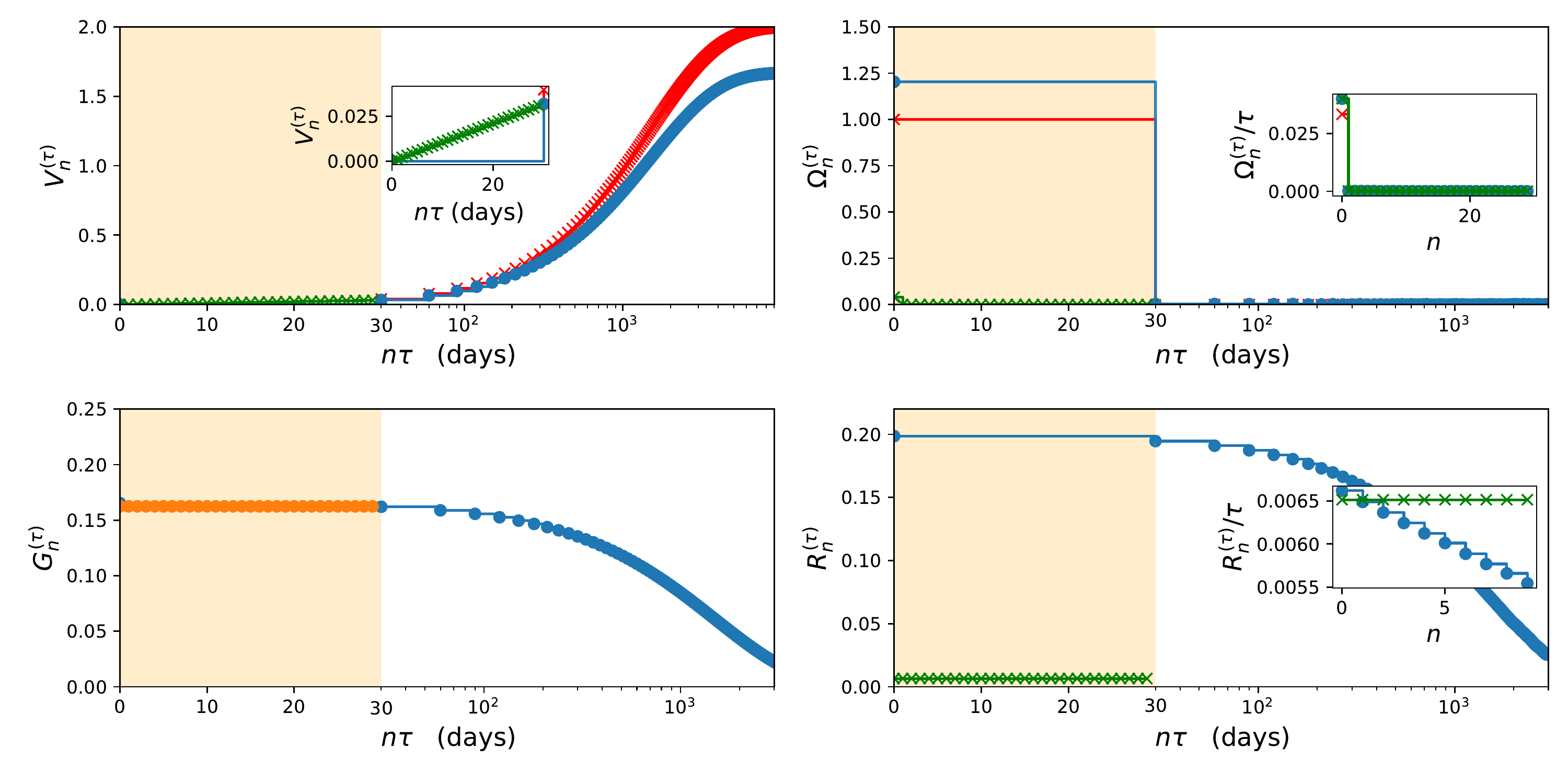}
    \caption{Independent order flow. Informed trader's estimate of the fundamental price are Markovian, with mean-reversion time scale $\tau_{\F} = 50$ days. At sampling scale $\tau_{\Long}$, the variance of the IT's estimate of the fundamental price is fixed to one, as well as the variance of the NT's order flow.}
    \label{fig:indep_sketch}
    \end{subfigure}
    
    \vspace{1cm}
    \begin{subfigure}{1\textwidth}
    \centering
    \includegraphics[scale = 0.5]{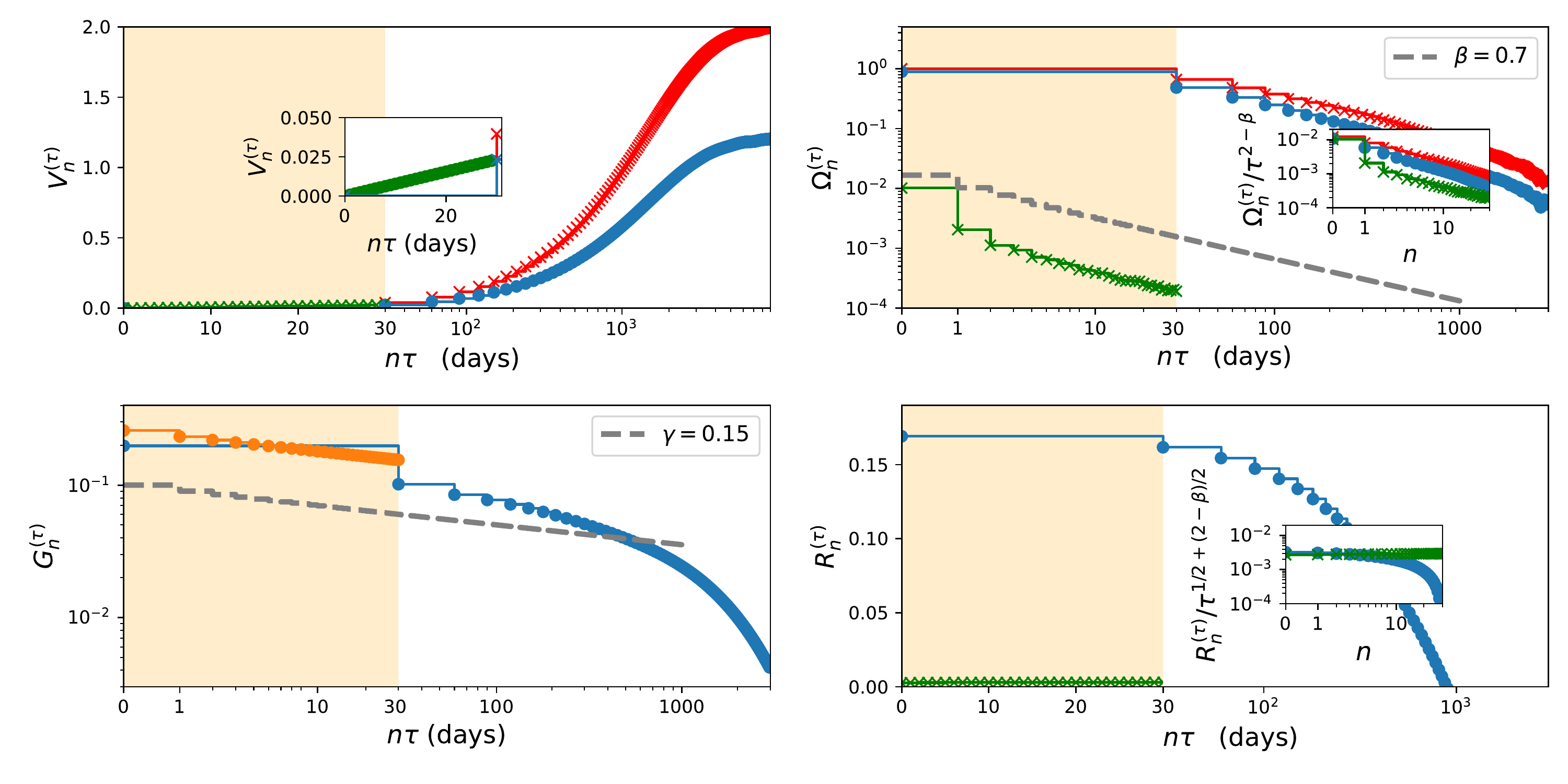}
    \caption{ Power law order flow. Informed trader's estimates of the fundamental price  are Markovian, with mean-reversion time scale $\tau_{\F} = 50$ days
    and NT's order flow ACF is a combination of exponentials that mimics a decreasing power law function with exponent $\beta = 0.7$. At sampling scale $\tau_{\Long}$, the variance of the IT's estimate of the fundamental price is fixed to one, as well as the variance of the NT's order flow. Dashed grey lines in top-right and bottom-left panels refer to decreasing power law functions with exponent $\beta$ and $\gamma$, respectively.  }
    \label{fig:pl_sketch}
    \end{subfigure}
    \caption{Calibrations related to cases (a) and (b) mentioned in Subsec.~\ref{sub:freq}, on  synthetic datasets. For each case, we consider two models with different sampling scale, i.e. $\tau$ equal to $\tau_{\short}=1$ day and $\tau_{\Long}=30$ days. We identify the high-frequency regime as the interval $[0,\tau_{\Long}]$ (orange band). The set of input ACFs that specify the sK model at sampling scale $\tau_{\Long}$ are related to NT's trades and to IT's fundamental price estimates (red lines). Details about them are given in  sub-captions. Once the sK model is solved, we obtain $ V_n^{(\tau_{\Long})}, \Omega_n^{(\tau_{\Long})}, G_n^{(\tau_{\Long})}$ and $R_n^{(\tau_{\Long})}$ (blue lines).
    The model with sampling scale $\tau_{\short}$, is calibrated with new input ACFs related to price and excess demand, and with the associated response function  (green lines). Note that in this case the response function is not an output of the model and so it is again presented as a green line. These new ACFs are such that, after a proper rescaling (see Subsec.~\ref{sec:multi}), the variogram, the excess demand ACF and the response function match the one at low frequency, as shown in insets. Results  of calibrations with sampling scale $\tau_{\short}$ are given by price impact  functions (orange lines), i.e., $G_n^{(\tau_{\short})}$.  }
    \label{fig:sketch}
    \end{figure}

Here, analyzing synthetic data, we provide graphical details about cases (a) and (b) mentioned in Subsec.~\ref{sub:freq}. Two calibrations with different sampling scales ($\tau_{\short}$ and $\tau_{\Long}$) are analyzed for each case. The calibration with $\tau_{\Long}$ is done on a time window that encompasses the low-frequency regime. Because of this, fundamental price mean reversion needs to be taken into account and we calibrate the sK model (with the numerical scheme detailed in Ref.~\cite{Vodret_2021}) starting from NT's order flow $q^\NT$ and IT's fundamental price estimate $p^\IT$ ACFs. Conversely, the calibration with sampling scale $\tau_{\short}$ is restricted only to the high-frequency regime (mimicking what is done usually when analyzing empirical data at high frequency). In this case, the numerical scheme used to solve the sK model cannot be applied because the model is not properly regularized, because ACFs do not have enough time to decay to zero. Because of this, we calibrate the model defined at  sampling scale $\tau_{\short}$ with the technique usually employed in the propagator model literature (details about it are given in App.~\ref{sub:high_freq_calib}).   Figure \ref{fig:sketch} shows that if the price and the order flow processes obtained at scale $\tau_{\Long}$ are compatible with those defined at sampling scale $\tau_{\short}$, the results of the two calibrations are compatible (via the coarse-graining argument explained in Subsec.~\ref{sec:multi}). In particular, the price impact functions calibrated with the two different sampling scales are compatible (see the bottom-left panels in Fig.~\ref{fig:sketch}). As a consistency check, note that calibrations related to case (b), where excess demand ACF's shape is a power law, validate the well-known constraint between the exponents of the power laws related to the excess demand ACF~($\beta$) and to the price impact functions ($\gamma$), given by $\gamma = (1-\beta)/2$ \cite{Propagator}. 

\section{Datasets, detrending and calibration procedures}
\subsection{Low frequency}
\label{sub:low_freq_calib}
\paragraph{Presentation of the data}
The dataset used for the calibration is about the S$\&$P500 index and is publicly available online at github.com/datasets/s-and-p-500.
These data  include information about monthly ($\tau =1$ month)  prices ($P^\emp_n$) and dividends ($M_n$) from January 1871 until March 2018, but do not include information about order flows. 

\begin{figure}
    \centering
    \includegraphics[scale = 0.52]{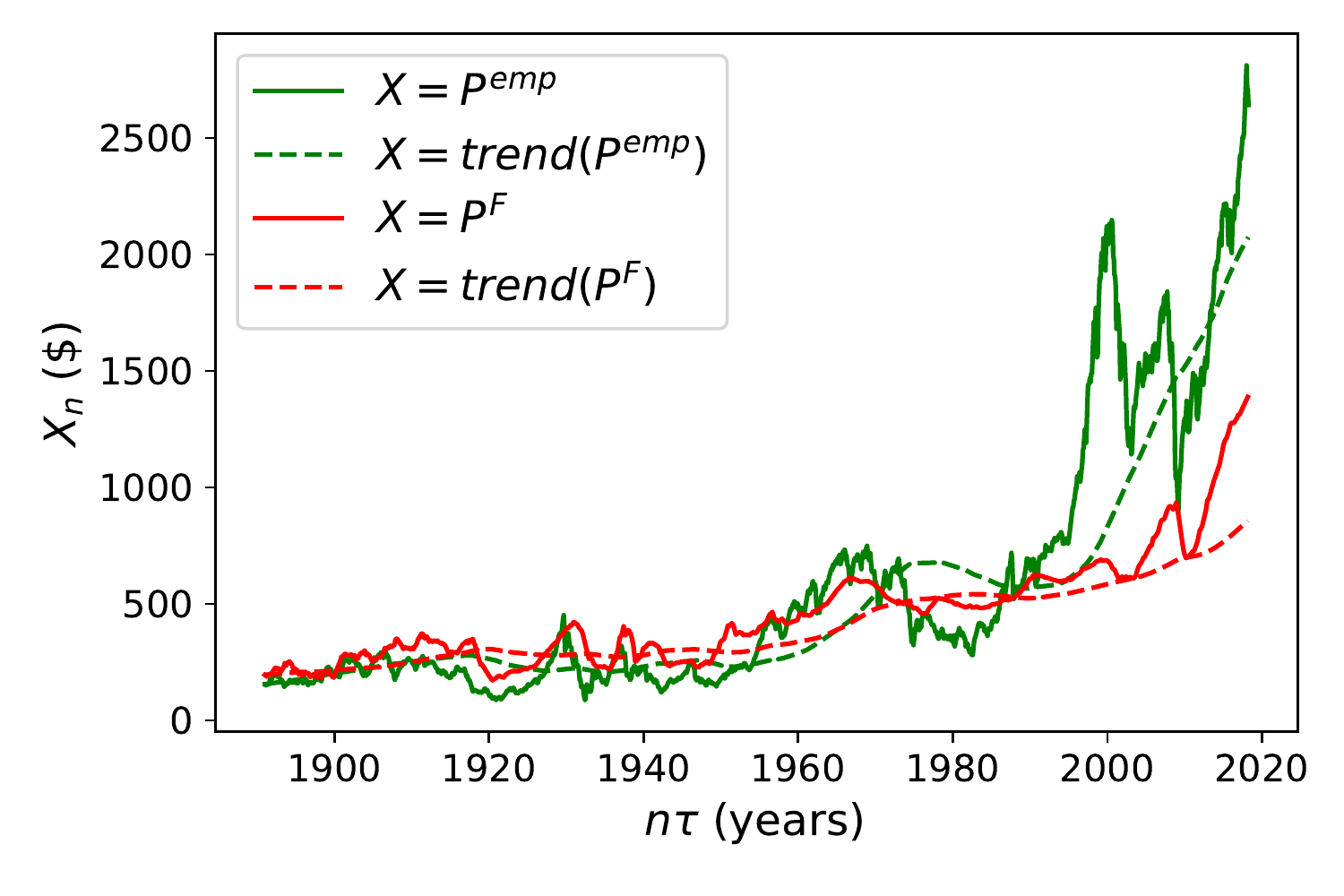}
    \includegraphics[scale = 0.52]{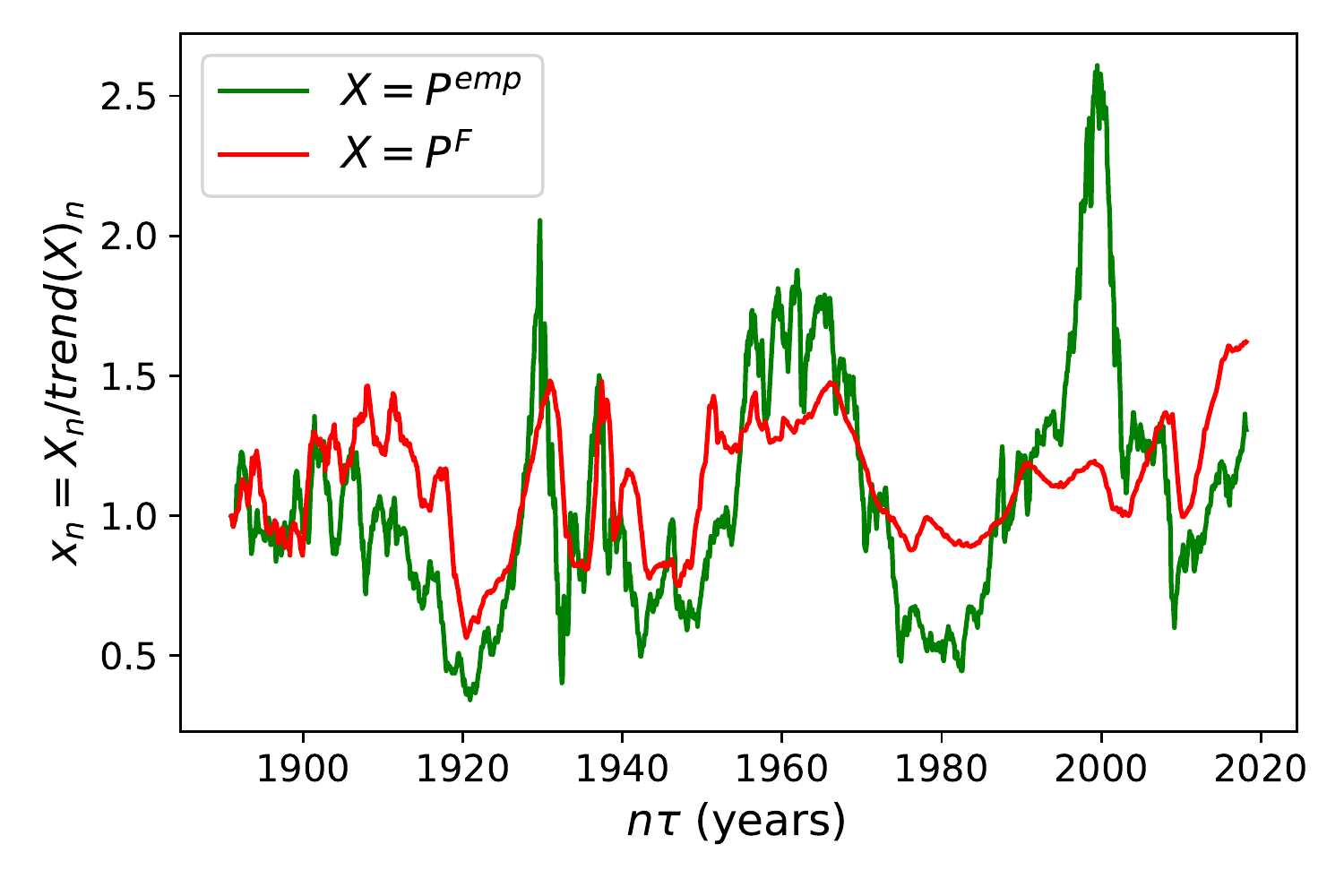}
    \caption{De-trending procedure applied to  S$\&$P500 index data. (Left) Empirical price and estimated fundamental price. In the legend, $trend(X)$ refers to the inverse of the exponential factor in Eq.~\ref{eq:param_de-trending}. (Right)
    Stationary version of empirical price and estimated fundamental price.}
    \label{fig:sep500}
\end{figure}

From data about dividends,
an estimate of the fundamental price can be constructed as $P^{\F}_n = M_n \langle P^\emp_n/M_n \rangle$\footnote{Let us note that this prescription for the fundamental price is not causal since the mean price-dividend ratio $\langle P_n/M_n\rangle$ is calculated with all the data provided by the dataset.}.
The price and the fundamental price cannot be described by a stationary process, as one can see from the left panel of Figure~\ref{fig:sep500}: a clear trend is exhibited by both  processes. Thus, we cannot calibrate the stationary Kyle model on these raw data. One needs to de-trend them. The de-trending procedure's goal is to retrieve a stationary empirical price and fundamental price processes, so that the calibration of the sK model, which is a stationary model, becomes possible.

We define de-trended quantities as:
\begin{equation}
\label{eq:param_de-trending}
    x_n = \frac{X_n}{X_0} \text{exp}\left(-\sum_{m=0}^n \eta^X_{m} \right), 
\end{equation}
where the trend $\eta^X_n>0$ is estimated in a causal way as follows. 

\paragraph{De-trending}

For each time series $X_n$ we define the trend in a causal way, as follows:
\begin{equation}
\label{eq: trend}
    \eta^X_n = \frac{1}{T/\tau} \log\left(\frac{X_n}{X_{n-T/\tau}}\right),
\end{equation}
where we choose $T = 20$ years in order to be able to capture phenomena  which occur on  time scales as large as few years, such as the mean-reversion of the fundamental price.

The  right panel of Figure~\ref{fig:sep500} shows the de-trended version of price and fundamental price processes calculated using Eqs.\eqref{eq:param_de-trending} and \eqref{eq: trend}. From the right panel of Fig.~\ref{fig:sep500} one can see that the de-trended fundamental price errs less than the de-trended price. This is the cause of price  excess-volatility, as emphasized in the main text.

Information about order flows is necessary to obtain a complete calibration, but in the dataset we are working with, no information is given about it. As we shall see in what follows, we can still partially calibrate the model, and obtain a complete description of the price process.

\paragraph{Calibration via the stationary Kyle model}

In this section we detail the calibration procedure, given the de-trended time series we obtained above. Firstly, we remove the mean price level from the empirical de-trended price, assuming that it is common knowledge so that price impact it's insensitive to it.

Let us now detail how we can completely characterize the price process without information about the excess demand.
Suppose that the signal and the noise ACFs are given by:
\begin{subeqnarray}
\label{eq:assumptions}
\Sigma^{\IT}_n&=& \Sigma^{\IT}_0 \alpha^n
\slabel{eq:assumptions_p_F}\\
\Omega^{\NT}_n&=& \Omega^{\NT}_0\delta_n \slabel{eq:NT_non_corr}
\end{subeqnarray}
where $0<\alpha<1$ and $\delta_n$ is the Kronecker's delta.
The first equation states that the estimate of the IT of the (de-trended) fundamental price follows an auto-regressive process of order 1, which is a first approximation widely used in the literature~(see, e.g.,  \cite{Blackok}). The second assumption states that on the scale of months the NT's order flow is not correlated, which is a good first approximation at very low frequency.

We have been able to characterize analytically the stationary equilibrium that arise within the stationary Kyle model in this case \cite{Vodret_2021}. 
Interestingly, in the linear stationary equilibrium, the variance of the price set by the MM is given by:
\begin{equation}
\label{eq:price_ratio}
    \frac{\Sigma^\SK_0}{\Sigma^{\IT}_0} =  \frac{1-\sqrt{1-\alpha^2}}{\alpha^2},
\end{equation}
which interpolates between the outcome of the original Kyle model  if the signal becomes very short-lived ($\alpha \rightarrow\\
 0$) and the case of fully revealing price if the signal becomes permanent ($\alpha \rightarrow 1$).
 Moreover,  in this case, information about the excess demand is not needed to specify completely the price process. 
 
 The calibration procedure goes as follows: from the fundamental price shown in the right panel (red line) of  Fig.~\ref{fig:sep500} one can calculate the associated empirical ACF. Then, we fit this empirical ACF with an exponential function, as prescribed by Eq.~\eqref{eq:assumptions_p_F}.  Although it's not possible to fix the amplitude of the price impact function $G^\SK$ because we lack information about the excess demand's variance $\Omega_0$, it's still possible to completely specify the price process by Eqs.~\eqref{eq:price_efficiency}, \eqref{eq:assumptions_p_F} and \eqref{eq:price_ratio}. 
 The results of this calibration are reported in Table \ref{tab:time scales}.

\subsection{High frequency}
\label{sub:high_freq_calib}

\paragraph{Presentation of the data}
We analyzed data  about some of the most traded stocks, during the period January 2013 - December 2020.

The tick size of all the stocks is 0.01 USD. We  reshaped  the data removing effects coming from stock splitting. The bid-ask spread of a large tick stock is most of the times equal to one tick, whereas small tick stocks have spreads that are typically a few ticks. There exist also a number of stocks in the intermediate region between large and small tick stocks, which have the characteristics of both types. We choose 5 different stocks and we placed them all  in the same pool. 
Data are indexed by a time label with precision at the microsecond ($\tau= 1 \mu s$), where information about the precise timing of the transaction is stored.

The empirical mid price $P^\emp_n$ is calculated as the mean between the bid and the ask.  At the transaction level, we constructed order-signs by labelling trades for which the transaction price is above the mid-price by $\epsilon_n = +1$ and all trade below as $\epsilon_n = -1$. Trades exactly at the mid-price were discarded. Signed order flow $q_n$ are then constructed multiplying the trade's sign by the quantity traded. Data about volumes are normalized by a rolling mean of the total daily volume exchanged over  a time window $\sim 50$ days.

\begin{figure}
    \centering
    \includegraphics[scale = 0.5]{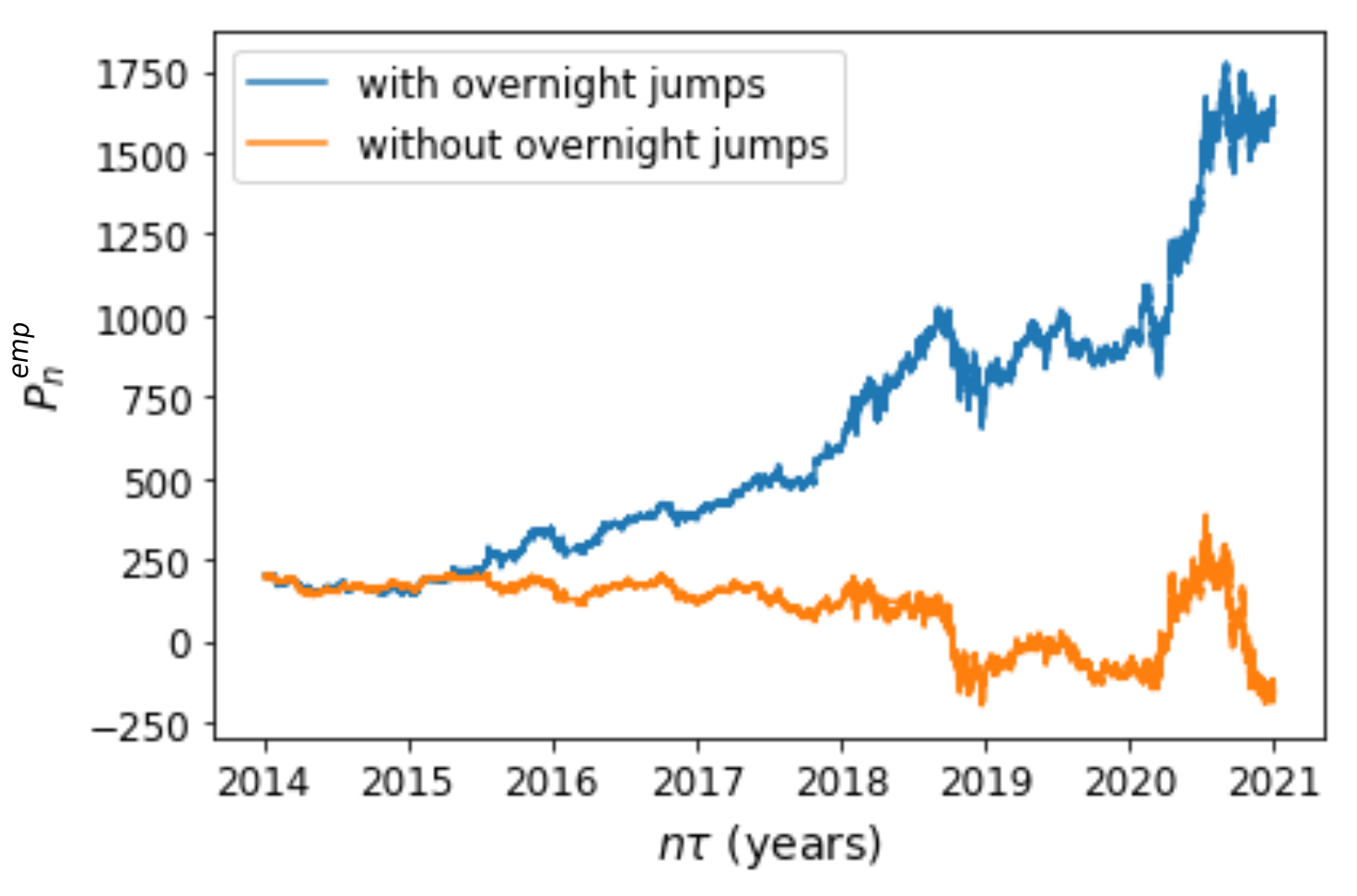}
    \includegraphics[scale = 0.5]{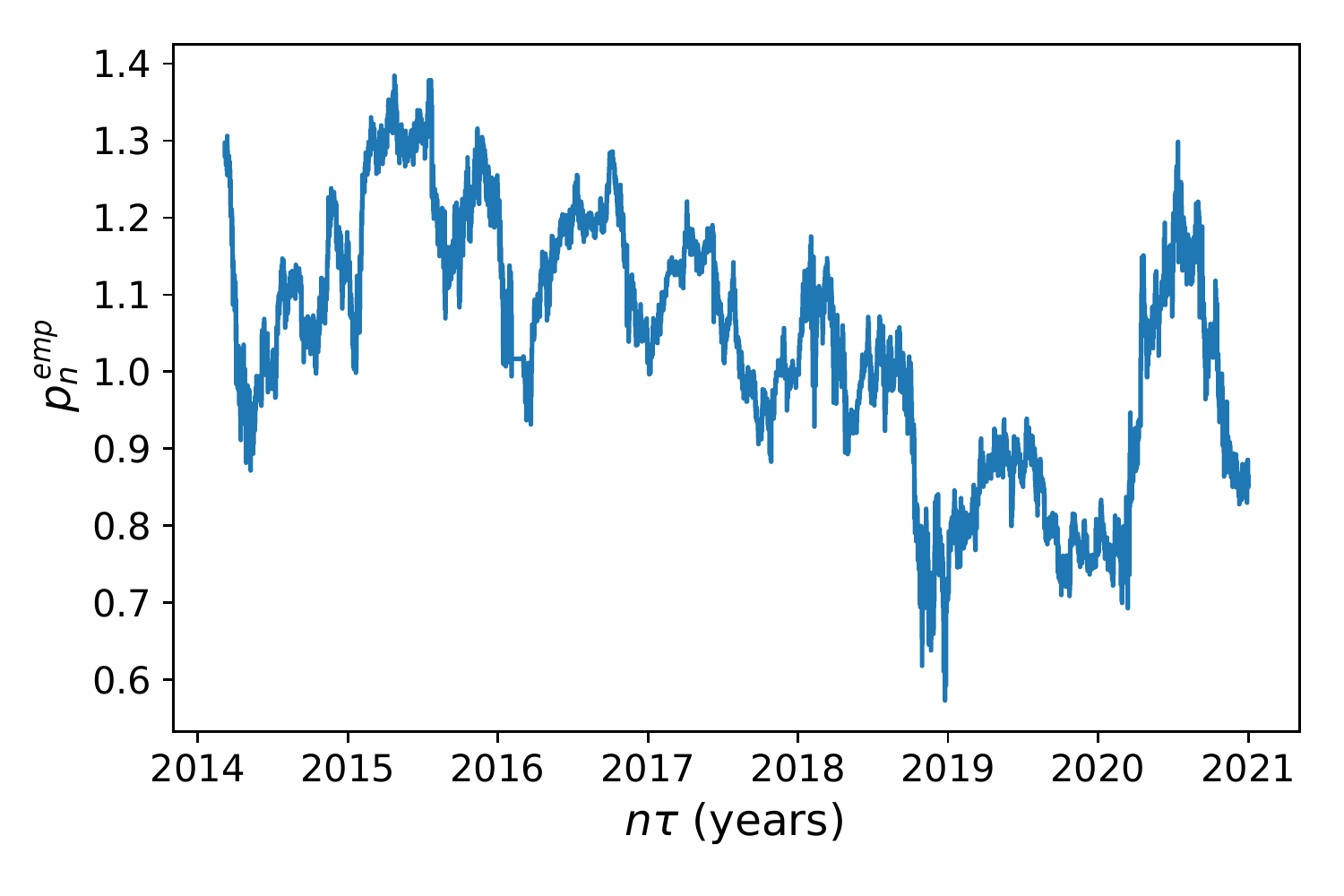}
    \caption{High frequency empirical data about AMZN stock. The sampling scale is equal to $\tau = 1\mu s$. (Left)~The raw empirical price is shown (blue line). The trend in the mean of the raw empirical price is removed if we remove  overnight jumps (orange line). (Right) We show the mid after the multiplicative de-trending procedure. The de-trended price doesn't exhibit trend nor in the mean, nor in the volatility.}
    \label{fig:my_de-trended}
\end{figure}

The empirical price exhibits a positive trend (as one can see from the left panel of  Figure \ref{fig:my_de-trended}), thus a de-trending procedure has to be implemented in order to meet the assumptions on which the stationary Kyle model is constructed. 
The de-trending procedure is discussed in detail below.

\paragraph{De-trending data at high frequency}
\label{app:high}

In order to proceed further, let us note that the main contribution to the trend in the price level is realized during the overnight, as the left panel of  Figure~\ref{fig:my_de-trended} shows. Removing overnight jumps is not enough to make the price stationary. In fact, as one can see from the orange line in the left panel of Fig.~\ref{fig:my_de-trended}, we still have a trend in volatility. In order to deal with this, we start from the price with overnights and we apply a logarithmic transformation. In doing this we obtain:
\begin{equation}
    \log\left(P^\emp_n \right) = \log({P^\emp_0})+\sum_{m=0}^n \eta_{m} + 
    \log(p^\emp_n),
\end{equation}
where $p^\emp_n$ is the residual part of the price, after the trend $\eta_n$ is removed.
After this logarithmic transformation, we remove the overnight jumps, removing -effectively- the trending component which depends on $\eta_n$. We do this by considering the trading activity in the period 9:30-16:00 in all  days under analysis. For each stock we concatenate the data on different trading days.
Then, we apply an exponential transformation in order to get back to the de-trended price. 
The result of this operation on the AMZN stock is shown in the right panel of  Figure~\ref{fig:my_de-trended}.

\paragraph{Calibration via propagator model}
\label{app:calib}
 In what follows we detail the calibration procedure for the linear version of the propagator model, defined by Eq.~\eqref{eq:propagator_model}.

From data about excess demands $q_n$ and returns $r^\emp_n = p^\emp_{n+1}-p^\emp_n$ we can measure the empirical return-response function $S^{\emp}_n = \langle q_m r^\emp_{n+m} \rangle$ and the empirical order flow ACF $\Omega_n$.
These two functions are related through:
\begin{equation}
    S^\emp_m=\sum_{n\geq 0} G^\Prop_n \Omega_{n-m}.
\end{equation}
The equation above can be inverted in order to obtain the price impact function $G^\Prop_n$. The relation between the response function $R^\emp_n$ and the return-response function is given by:
\begin{equation}
    R^\emp_n = \sum_{0 \leq m < n} S^\emp_m
\end{equation}
allowing to recover the response function from its differential form.
\end{document}